\documentclass[11pt,letterpaper]{article}
\usepackage[T1]{fontenc}
\usepackage[margin=1in]{geometry}
\usepackage{appendix}
\let\textcaronaccent\v
\usepackage{latex_helpers/preamble}

\usepackage[
  backend=biber,
  style=alphabetic,
  natbib=true,
  maxbibnames=99,
  maxcitenames=99,
  maxalphanames=4,
  minalphanames=3,
  date=year,
  isbn=false
]{biblatex}
\AtEveryBibitem{%
  \clearfield{urlyear}%
  \clearfield{eventtitle}%
  \iffieldundef{doi}{}{\clearfield{url}}%
}
\DeclareSourcemap{
  \maps[datatype=bibtex]{
    \map[overwrite]{
      \step[fieldsource=entrykey, match=\regexp{\ABKN23\z}, final]
      \step[typesource=article, typetarget=inproceedings]
      \step[fieldset=booktitle, fieldvalue={40th International Symposium on Theoretical Aspects of Computer Science ({STACS} 2023)}]
      \step[fieldset=series, fieldvalue={Leibniz International Proceedings in Informatics ({LIPIcs})}]
      \step[fieldset=journaltitle, null]
      \step[fieldset=note, null]
      \step[fieldset=doi, fieldvalue={10.4230/LIPIcs.STACS.2023.11}]
    }
    \map{
      \step[fieldsource=entrykey, match=\regexp{\AMP13\z}, final]
      \step[fieldsource=author, fieldtarget=editor]
      \step[fieldset=edition, null]
    }
  }
}

\usepackage{enumitem}
\AtBeginBibliography{\let\v\textcaronaccent}
\usepackage{booktabs}

\usepackage{mathtools}
\usepackage{microtype}
\usepackage{tabularx}
\usepackage{needspace}
\usetikzlibrary{arrows.meta,decorations.pathreplacing,backgrounds}

\newcommand{\ZetaTransform}{\mathsf{Z}}
\newcommand{\IncidenceTransform}{\mathsf{I}}
\newcommand{\AffineTransform}{\mathsf{A}}
\newcommand{\Gr}{\operatorname{Gr}_q}
\newcommand{\GL}{\operatorname{GL}}
\newcommand{\qi}[1]{[#1]_q}
\newcommand{\ETH}{\mathrm{ETH}}

\crefname{thm}{theorem}{theorems}
\Crefname{thm}{Theorem}{Theorems}
\crefname{lem}{lemma}{lemmas}
\Crefname{lem}{Lemma}{Lemmas}
\crefname{prop}{proposition}{propositions}
\Crefname{prop}{Proposition}{Propositions}
\crefname{cor}{corollary}{corollaries}
\Crefname{cor}{Corollary}{Corollaries}
\crefname{rem}{remark}{remarks}
\Crefname{rem}{Remark}{Remarks}
\crefname{example}{example}{examples}
\Crefname{example}{Example}{Examples}

\setlist{nosep,leftmargin=1.5em}
\DontPrintSemicolon
\SetKw{Return}{return}
\SetKw{Continue}{continue}
\SetAlFnt{\normalsize}
\SetAlCapFnt{\normalsize\bfseries}
\SetAlCapNameFnt{\normalsize\bfseries}
\SetAlgoSkip{medskip}
\renewenvironment{abstract}
  {\begin{center}\normalfont\normalsize\bfseries\abstractname\end{center}
   \begin{quotation}\normalfont\normalsize}
  {\end{quotation}}

\title{A Fast Algorithm for Subspace Convolution}
\author{
Shlomo A. Perles\thanks{\raggedright Research supported by ERC Starting grant
CODY 101039914, and ISF grant 3011005535.
\texttt{shlomo.perles@mail.huji.ac.il}}\\
 The Hebrew University of Jerusalem\\
  \and
  Omri Weinstein\thanks{\raggedright Supported by an ERC Starting grant CODY
101039914, and ISF grant 3011005535, \texttt{omriwe@cs.huji.ac.il}}\\
  The Hebrew University of Jerusalem}
\date{}

\hypersetup{pdftitle={A Fast Algorithm for Subspace Convolution},
  pdfauthor={Shlomo A. Perles and Omri Weinstein}}

\begin{document}
\pagenumbering{roman}
\maketitle
\thispagestyle{empty}

\begin{abstract}
A cornerstone result in Boolean lattice theory is the subset-zeta transform, which
computes the $f$-weighted subset summation problem $(Z f)(B)=\sum_{A\subseteq B}f(A)$ for all $ B\subseteq[n]$
in $O(N \log N)$ additions instead of the na\"ive $O(3^n) = O(N^{\log_2 3})$, where $N=2^n$.

This paper studies the algebraic analogue of this problem--the \emph{subspace summation} problem--where subsets are replaced by linear \emph{subspaces} in $\mathbb{F}_q^n$. Surprisingly, we show that this problem admits a similar $\Theta_q(N_q \log N_q)$-time algorithm, where $N_q$ is the number of subspaces. This improves substantially (in both theory and practice) over lattice-based algorithms (\cite{BHK+16,KKW16}), which are known to be optimal for addition-only algorithms over $\mathbb R$.
We prove similar runtimes for inversion, rank-constrained sums and span convolution. 
We describe two main applications:

\begin{itemize}
    \item \textbf{Fast linear algebra on Grassmann graphs:} Matrix--vector multiplication for arbitrary distance-invariant kernels on $N$-vertex Grassmann graphs in $O_q(N\log N)$ arithmetic operations, together with fast spectral projections and exact solutions of the corresponding linear systems.
    \item \textbf{Graph coloring of low binary cut-rank graphs:} An $n^{2^{r^2/4+O(1)}}$-time algorithm for computing the chromatic number and chromatic polynomial of an $n$-vertex graph, given a rank-decomposition of width $r$. We also give quantitative reductions from satisfiability and monotone constraint problems.

\end{itemize}
Experiments show speedups in measured running time of up to $18$ over the
covering-pair algorithm for the zeta transform, $190$ over Poisson summation
for rank-constrained sums, and $22$ over the distance-matrix recurrence
for Grassmann kernel multiplication.
Lean~4 formalizations of the transform identities, arithmetic operation
bounds, and graph-coloring constructions accompany the paper~\cite{PW26code}.
\end{abstract}
\clearpage
\pagenumbering{arabic}

\section{Introduction}
A fundamental problem in Boolean lattice theory is \emph{subset summation}: given a coefficient $f(A)\in\mathbb{R}$ for every $A\subseteq[n]$,
compute $(\ZetaTransform f)(B)=\sum_{A\subseteq B}f(A)$ for every
$B\subseteq[n]$, where $f,\ZetaTransform f\in\mathbb{R}^N$,
and $N=2^n$. For a given subset $B$, the sum has $2^{|B|}$ terms
and takes at most $N-1$ additions. Computing all sums separately
enumerates $\sum_B2^{|B|}=3^n = N^{\log_2 3}$ terms.
The classical \emph{fast zeta transform} (see~\cite[Section~1.1]{KKW16}) uses only $O(n2^n)=O(N\log N)$
\emph{additions} to evaluate all sums. A similar algorithm applies to the inverse subset-sum problem, via the M\"obius transform~\cite{KKW16}, which is the inverse of the zeta transform. These transforms led to profound algorithmic speedups for a variety of combinatorial problems, including the
Chromatic Number, Hamiltonian path and TSP problems \cite{BHK09, BHKK07, Bjo10}, Steiner tree decompositions \cite{BHKK07}, perfect matchings and cycle covers \cite{BHK09}.

We study the analogous problem for \emph{linear
subspaces}. Throughout, $\F_q$ is a finite field of fixed order
$q=p^m$, where $p$ is prime, and $\Gr(n)$ is the set of all subspaces
of $\F_q^n$. Given $f:\Gr(n)\to\mathcal R$, where $\mathcal R$ is a
commutative coefficient ring, compute
\begin{equation}\label{eq:zeta}
  (\ZetaTransform_n f)(W)=\sum_{U\le W}f(U),\qquad W\in\Gr(n).
\end{equation}
Here $U\le W$ means that $U$ is a linear subspace of $W$. Thus
$f$ and $\ZetaTransform_n f$ each have $N_q(n)=|\Gr(n)|$ entries.
The matrix $\ZetaTransform_n$ has entry $\ZetaTransform_n(W,U)=1$
if $U\le W$ and zero otherwise. For example, over $\Q$, taking $f$
to be the indicator of a family of subspaces makes $(\ZetaTransform_n f)(W)$
count the members of that family contained in $W$.
The inverse of $\ZetaTransform_n$ is called the M\"obius transform,
and the transpose $\ZetaTransform_n^{\mathsf T}$ sums over superspaces.

The subset algorithm passes information from each set to its immediate
supersets, of which there are at most $n$. A subspace $U\le\F_q^n$
has $(q^{n-\dim U}-1)/(q-1)$ immediate superspaces, making the same
approach much more expensive --
the number of subspaces in $\F_q^n$ is
$N_q(n)=\Theta_q(q^{\lfloor n^2/4\rfloor})$ (see Appendix~\ref{sec:counts}), whereas the number of
\emph{covering pairs} $U<W$ with $\dim W=\dim U+1$ is
$\Theta_q(N_q(n)q^{n/2})$ (see \eqref{eq:cover-pairs}). The lattice algorithms
of~\cite[Theorem~1.3]{KKW16, BHK+16} therefore take $\Theta_q(N_q(n)q^{n/2})=\Theta_q\left(N_q(n)^{1+1/\sqrt{\log_q N_q(n)}}\right)$ arithmetic operations, proportional to the number of these pairs.
We call this approach the \emph{covering-pair algorithm}.
Over $\mathbb R$, Kennes~\cite[Theorem~6]{KenMarch-April/1992} proves a matching lower bound for algorithms using only
additions.

Our algorithm avoids this addition-only lower bound by using subtraction
and scalar multiplication. The resulting cancellations let us sum over
contained subspaces without listing all these pairs. The core idea behind our \emph{fast subspace transform} is a recursive hyperplane cutting technique, following Srinivasan's character decomposition~\cite{Sri14}. This reduces affine containment of \emph{crossing subspaces} to a translation-invariant sum, allowing an ordinary FFT to decompose the subspace sum problem into lower-dimensional subspace summation problems in the frequency domain.

\subsection{Our Results}\label{sec:results}
Let $\Gr(n,j)$ be the set of all $j$-dimensional subspaces in $\F_q^n$.
The number of subspaces of dimension at most $k$ is
\[
 M_q(n,k)=\sum_{j=0}^{\min(k,n)}|\Gr(n,j)|,\qquad M_q(n,n)=N_q(n).
\]
The truncated containment matrix $\ZetaTransform_{n,k}$ is the
$M_q(n,k)\times M_q(n,k)$ matrix with entries $\ind[U\le W]$,
where $\ind[\cdot]$ is one when its condition holds and zero otherwise,
and both indices have dimension at most $k$. Ordering by dimension makes it triangular with diagonal one. For $k=n$, this is
the full transform $\ZetaTransform_n$. Its transpose sums over superspaces
in the same table. Our algorithms require the following assumption
on the coefficient ring $\mathcal R$.

\begin{thm}[Fast Subspace Summation]\label{thm:linear}
Over any commutative ring in which $p$ is invertible, each of
$\ZetaTransform_{n,k}$, $\ZetaTransform_{n,k}^{-1}$, $\ZetaTransform_{n,k}^{\mathsf T}$, and
$(\ZetaTransform_{n,k}^{\mathsf T})^{-1}$ can be applied in
$O_q(nkM_q(n,k))$ arithmetic operations for $k\le n$.
In particular, the cost is $O_q(M_q(n,k)\log M_q(n,k))$.
\end{thm}

\paragraph{Corollary: Fast Subspace Convolution}
Zeta transforms also turn convolution into pointwise multiplication.
For two subspace coefficient functions $f,g$ on $\Gr(n)$, define their \emph{span convolution} by
\[
 (f\star g)(W)=\sum_{U+V=W}f(U)g(V),
\]
where the sum is over ordered pairs of subspaces. Since
$U+V\le W$ exactly when $U,V\le W$,
$\ZetaTransform_n(f\star g)=(\ZetaTransform_n f)(\ZetaTransform_n g)$,
with pointwise multiplication on the right. By \cref{thm:linear},
two transforms and one inverse compute $f\star g$ in
$O_q(N_q(n)\log N_q(n))$ arithmetic operations.

\paragraph{Low-Rank Summation}
One may instead have weights only on $a$-dimensional subspaces
and want sums only on $b$-dimensional subspaces. For example, given
weights on one-dimensional subspaces (projective points), compute
their sums in every hyperplane. In this example, the input and output tables have
$q^{O(n)}$ entries, whereas the full subspace table has
$q^{\Theta(n^2)}$ entries. Consequently, applying the full transform takes
$q^{\Theta(n^2)}$ work, whereas \cref{thm:incidence} below brings this
down to $q^{O(n)}$.
For a coefficient ring $\mathcal R$ and
$0\le a\le b\le n$, define the \emph{incidence transform} by
\[
 (\IncidenceTransform_n^{a,b}f)(W)=\sum_{\substack{U\le W\\\dim U=a}}f(U),
 \qquad f:\Gr(n,a)\to\mathcal R,\quad W\in\Gr(n,b).
\]
Its matrix has $|\Gr(n,b)|$ rows
and $|\Gr(n,a)|$ columns. The transpose sends weights on
$b$-dimensional subspaces to sums over those containing each
$a$-dimensional subspace.
\begin{thm}\label{thm:incidence}
Under the coefficient assumption of \cref{thm:linear}, $\IncidenceTransform_n^{a,b}$
and its transpose take $O_q(S\log S)$ arithmetic operations,
where $S=|\Gr(n,a)|+|\Gr(n,b)|$ is the total input and output length.
\end{thm}
For $(n,a,b)=(4t,t,3t)$, this bound is optimal in the bounded-coefficient
model over $\C$ (see Appendix~\ref{sec:lower-bound}).
Constructing the subspace bases and coordinate changes used to index
these tables takes table size times $\poly(n)$ additional work.
Appendix~\ref{app:implementation} gives the construction and exact bounds.

\subsubsection{Applications}

\paragraph{Fast linear algebra on Grassmann graphs}

A \emph{geometric kernel} assigns a weight $K(x_i,x_j)$ to each pair
of points $x_i \in \R^\ell$.
A standard example is the radial kernel $K(x,y)=f(\|x-y\|_2^2)$.
For $N$ points $x_i \in \R^\ell$, $K$ implicitly defines an $N\times N$ matrix using only $N\ell$ coordinates and the function $f$. If each coordinate uses $\poly(\log(N\ell))$ bits and $f$ has a bounded-size description, this is $\widetilde O(N\ell)$ information bits. As such, one could hope to \emph{multiply} $K$ by a vector (or perform other linear-algebraic operations)
in subquadratic time ($\ll N^2$). The celebrated Fast Multipole Method of Greengard and Rokhlin~%
\cite{GR87} %
gives such approximate matrix--vector products for suitable kernels. For Gaussian kernels, Greengard and Strain developed the Fast Gauss Transform~%
\cite{GS91}%
. Alman, Chu, Schild, and Song~\cite{ACSS20}
initiate a systematic study of fast linear algebra on geometric
graphs, including approximate matrix--vector multiplication,
spectral sparsification, and Laplacian solving.

 We study a finite-geometric analogue: kernels determined by
 intersection dimensions on the full Grassmannian.
Let $V=\Gr(n,k)$, $N=|V|$, and $d=\min(k,n-k)$.
The Grassmann graph $J_q(n,k)$ has vertex set $V$, where two vertices
$U,W$ are adjacent exactly when $\dim(U\cap W)=k-1$.
Their graph distance is $\operatorname{dist}(U,W)=k-\dim(U\cap W)$,
and its maximum is $d$.
An \emph{intersection kernel} is an $N\times N$ matrix specified by
coefficients $a_0,\ldots,a_d\in\mathcal R$.\footnote{Embed $U$ as $x_U=(\ind[L\le U])_{L\in\Gr(n,1)}$.
Then $\|x_U-x_W\|_2^2=2(q^k-q^{\dim(U\cap W)})/(q-1)$,
so intersection kernels are radial kernels on these vectors.}
Given weights
$f:V\to\mathcal R$, we want to compute
 \[
 K(U,W)=a_{\operatorname{dist}(U,W)},\qquad
 (Kf)(U)=\sum_{W\in V}a_{\operatorname{dist}(U,W)}f(W).
\]
\Needspace{6\baselineskip}
Our fast subspace transform computes this product without
constructing all $N^2$ entries:

\begin{thm}[Fast multiplication on the Grassmann graph]\label{thm:grassmann-intro}\label{thm:kernels}
Under the coefficient assumption of \cref{thm:linear}, given
 $a_0,\ldots,a_d\in\mathcal R$ and $f:V\to\mathcal R$,
the product $Kf$ can be computed in $O_q(N\log N)$ arithmetic operations in $\mathcal R$.
\end{thm}

 Earlier covering-pair transforms~\cite{KKW16} give $O_q(q^dN)$ operations.
Our bound replaces $q^d$ by $\log N$, a substantial saving when
$q^d$ is much larger.

For example, choosing $a_j=\ind[j=i]$ gives the distance matrix $D_i$,
which sums weights at distance $i$:
$(D_if)(U)=\sum_{W\in V:\,\operatorname{dist}(U,W)=i}f(W)$.
In particular, $D_1$ is the adjacency matrix and sums neighbors' weights.
If $f$ is the indicator of a nonempty family $\mathcal C\subseteq V$,
computing over $\Q$ counts its members at each distance from $U$.
The smallest $i$ with $(D_if)(U)>0$ is the distance to its nearest member.

 We also compute all projections of $f$ onto eigenspaces of $D_1$
over $\Q$ or $\C$ in $O_q((d+1)N\log N)$ operations.
Over any field of characteristic different from $p$, the same bound
tests whether $K$ is nonsingular and, if so, solves $Kx=b$.
For the spectral task, Maslen, Orrison, and Rockmore~\cite[Theorem 5.8]{MOR03}
use $O(d[d]_qN)$ operations, where $[d]_q=(q^d-1)/(q-1)$.
Our bound improves the dependence on $d$ when $[d]_q$ is much
larger than $\log N$, though their bound can be smaller for small $d$.
Section~\ref{sec:applications} proves the multiplication bound, and
Appendix~\ref{app:grassmann} gives the distance sums and further applications.

\paragraph{Experiments.}
The constants hidden in our asymptotic bounds could make our algorithms
slower in practice, so we tested them against the comparison algorithms
in Table~\ref{tab:exp-summary}. Our measured running times show speedups of up
to $18$ for the zeta transform $\ZetaTransform_n$, $190$ for incidence
$\IncidenceTransform_n^{a,b}$, and $22$ for Grassmann kernel multiplication.
On these instances, the operation counts for the comparison algorithms
reach $4.1$, $170$, and $4.6$ times ours, respectively.
Both the zeta and incidence algorithms also handle instances
for which the comparison methods exceed the
available memory. These results are discussed in more detail in
Section~\ref{sec:experiments}.

\paragraph{Graph coloring of low binary cut-rank graphs}

Subspace containment also describes which color classes can share a
color across a graph cut. Our transform sums over colorings that can
be combined across the cut, and the same subspace representation leads to an improved
algorithm for coloring the whole graph.
For a graph $G$ with adjacency matrix $A_G$, the \emph{binary cut-rank}
of a cut is
\[
 \rho_G(X):=\operatorname{rank}_{\F_2}A_G[X,V(G)\setminus X].
\]
Across a cut of binary cut-rank $r$, vertices can be labeled by columns
$p_x,q_y\in\F_2^r$ so that $xy$ is an edge exactly when
$p_x^{\mathsf T}q_y=1$. A color class on either side is represented by
the \emph{span} of its labels, called its \emph{profile}. The number of profiles is at most the number of subspaces of $\F_2^r$, which is
\[N_2(r) := 2^{r^2/4+O(1)}\] 
(see Section~\ref{sec:coloring}). 
Two color classes on opposite sides can share a color precisely when their spans $U,V$ are orthogonal, or equivalently
$V\le U^\perp$. Given weights $g(V)$ for all $V\le\F_2^r$,
\cref{thm:linear} computes
$\sum_{V\le U^\perp}g(V)=(\ZetaTransform_r g)(U^\perp)$ for every $U$. Moreover, 
\cref{prop:color-compatibility} computes the corresponding sums for
several colors in time nearly linear in the number of input weights.

To color the whole graph, we need to account for the number of classes having each of the
$N_2(r)=2^{r^2/4+O(1)}$ possible subspace profiles. Our algorithm combines these counts without applying the transform.
A \emph{rank-decomposition} is a subcubic tree whose leaves are in
bijection with $V(G)$, and its width is the maximum binary cut-rank over
the cuts induced by deleting a tree edge. The \emph{rank-width}
$\operatorname{rw}(G)$ is the minimum width over all rank-decompositions.
A \emph{linear layout} is a vertex ordering $\pi=(v_1,\ldots,v_n)$,
whose width is
$r(\pi)=\max_{1\le i<n}\rho_G(\{v_1,\ldots,v_i\})$.
The \emph{linear rank-width} $\operatorname{lrw}(G)$ is the minimum
over these orderings. We assume that the decomposition or layout is supplied.

For a fixed number $c$ of colors, Bui-Xuan, Telle, and
Vatshelle~\cite{BTV10} give a $2^{O(cr^2)}n^{O(1)}$-time algorithm.
For unrestricted palettes, Ganian, Hlin\textcaronaccent{e}n\'y, and
Obdr\textcaronaccent{z}\'alek~\cite[Theorems~4.1 and~4.7]{GHO13}
compute the chromatic
number and chromatic polynomial in $n^{2^{r^2/2+O(r)}}$ time by
recording the multiplicities of subspace profiles and the ways to
pair color classes. We improve the exponent by processing these
pairings one profile at a time:

\begin{thm}[informal]\label{thm:coloring}
Given an $n$-vertex graph $G$ and a rank-decomposition of width $r$,
the chromatic number, an optimal coloring, and the chromatic polynomial
of $G$ can be computed in time
\[
 n^{O(N_2(r))}=n^{2^{r^2/4+O(1)}}.
\]
Assuming ETH~\cite{IP01}, the leading
coefficient $1/4$ is optimal, even for supplied linear layouts.
\end{thm}

The algorithm takes $n^{2^{r^2/4+O(1)}}$ time, whereas ETH rules out
even $n^{2^{r^2/4-17r}}$ time. See \cref{thm:general-coloring,xp:lower} for the precise bounds.\footnote{A recent preprint reports subexponential algorithms for deciding and counting $3$-SAT~\cite{Claude26ETH}. If correct, these refute ETH and counting ETH. Our reductions remain unconditional implications between running times, as explained in \cref{sec:overview}.}

For every fixed number $c\ge3$ of colors, ETH also excludes
$2^{o(r^2)}n^{O(1)}$ time, even with a supplied linear layout
(\cref{thm:fixed-palette}). Under this hypothesis, the bound matches the exponent order of the
algorithm of~\cite{BTV10} and answers the coloring question
in~\cite[Section~8, full version]{BKN23}.
Appendix~\ref{cnt:section} gives the corresponding fixed-color
counting lower bound under counting ETH.

\paragraph{Formal verification.}
The companion repository~\cite{PW26code} contains Lean~4 proofs of the
transform identities, arithmetic operation bounds, and graph-coloring
constructions, using Mathlib, together with an axiom audit.
The coloring lower bounds are formalized as conditional running-time implications.
Index construction, some bit-cost arguments, and the machine-model
running times of the reductions remain outside the formalization.
For the fixed-palette reduction, the formalization uses one clause-local
group per literal occurrence. The repository documents this variant,
the cost-model assumptions, and the remaining unformalized claims.

\subsection{Technical overview}\label{sec:overview}

This section provides a streamlined overview of the proof of Theorem~\ref{thm:linear}, its extensions, and the applications in Section~\ref{sec:results}.

\subsubsection{Fast zeta transform}
A natural approach to achieving a fast zeta transform is to partition $\Gr(n)$
into two sets and apply a smaller zeta transform to each. To make
this precise, call a set
$\mathcal{S}$ of subspaces an \emph{induced copy} of $\Gr(i)$ if there is a
bijection $\phi:\mathcal{S}\to\Gr(i)$ such that
\[
 U_1\le U_2\quad\Longleftrightarrow\quad\phi(U_1)\le\phi(U_2),
 \qquad U_1,U_2\in\mathcal{S}.
\]
If the two sets were induced copies of $\Gr(i)$ and $\Gr(j)$,
with $i,j<n$, and had no containments between them, ordering
the input and output coordinates by these sets would give
\[
 \ZetaTransform_n f=\operatorname{diag}(\ZetaTransform_i,\ZetaTransform_j)f.
\]
We could then apply the smaller transforms independently and repeat this
decomposition recursively. However, no partition into two nonempty
sets has this property.\footnote{The set containing the zero subspace
has a containment relation with every subspace in the other set.}

Instead, we seek a partition of $\Gr(n)$ in which one set is an
induced copy of a smaller subspace lattice on which
one smaller transform is immediate, and the other set requires extra care. Fix the hyperplane
$H=\F_q^{n-1}\times\{0\}$. The subspaces contained in $H$ form an
induced copy of $\Gr(n-1)$, so we can apply $\ZetaTransform_{n-1}$ there.
The remaining subspaces are the \emph{crossing subspaces}, whose set is
\[
 \mathcal X_H=\{U\le\F_q^n:U\not\le H\}.
\]
Let $f_0$ and $f_1$ be the restrictions of $f$ to the subspaces inside
$H$ and to $\mathcal X_H$, respectively. Ordering both rows and columns
with the inside subspaces first gives
\begin{equation}\label{eq:overview-cut}
 \ZetaTransform_n=
 \begin{pmatrix}\ZetaTransform_{n-1}&0\\ C&\AffineTransform\end{pmatrix},
 \qquad
 \ZetaTransform_n\binom{f_0}{f_1}
 =\binom{\ZetaTransform_{n-1}f_0}{Cf_0+\AffineTransform f_1}.
\end{equation}
The zero block appears because a crossing subspace cannot be contained
in an inside subspace. The block $C$ sums contributions from inside
subspaces to crossing subspaces, while $\AffineTransform$ sums over
containments between crossing subspaces.

For a crossing subspace $W$, the inside subspaces contained in $W$
are exactly the subspaces of $W\cap H$. Therefore
\[
 (Cf_0)(W)=\sum_{\substack{U\le W\\U\le H}}f(U)
          =(\ZetaTransform_{n-1}f_0)(W\cap H).
\]
This value is already available from the transform inside $H$, so
only $\AffineTransform f_1$ remains to be computed.

\paragraph{Splitting the crossing set.}
One may try partitioning $\mathcal X_H$ into induced copies of smaller
sets $\Gr(i)$, so that a zeta transform can be applied to each.
However, containments between these sets would still make their
transforms depend on each other. We therefore relax the induced-copy
requirement and allow each crossing subspace to contribute its
weight to several copies of smaller sets $\Gr(i)$. In each copy,
sum the contributions to the same subspace and apply its zeta transform.
Then recover the crossing outputs from all the transformed copies.
In matrix form, this asks for an invertible map $\mathcal F$ satisfying
\[
 \underbrace{\mathcal F(\AffineTransform f_1)}_{\text{copied outputs}}
 =\operatorname{diag}(\ZetaTransform_{i_1},\ldots,\ZetaTransform_{i_m})
  \underbrace{\mathcal F f_1}_{\text{copied inputs}},\qquad i_j<n.
\]
Summing contributions at the same subspace may lose information,
so we must choose the contributions so that $\mathcal F$ is reversible.
Fourier transformation is a natural choice: it is reversible and
can be computed quickly by FFTs.

\paragraph{Choosing the smaller copies.}
We use one copy of $\Gr(n-1)$ and $q^{n-1}-1$ copies of $\Gr(n-2)$,\footnote{Their combined size is
$N_q(n-1)+(q^{n-1}-1)N_q(n-2)=|\mathcal X_H|$, by
\eqref{eq:count} with cutoff $k=n$. For $n=1$, only the copy
labeled by $\lambda=0$ occurs.} labeling them by the frequencies $\lambda\in\F_q^{n-1}$. We regard each vector $\lambda$ as the linear functional
$\lambda(x)=\lambda^{\mathsf T}x$, so its copy consists of the subspaces
of $\ker\lambda$. This kernel has dimension $n-1$ when $\lambda=0$
and $n-2$ otherwise. Each frequency labels a separate copy, even
when two kernels coincide.

\paragraph{Choosing the Fourier transform.}
At frequency $\lambda$, applying Fourier transformation computes the contribution
of $f_1$ to the corresponding copy.
It remains to determine which specific subspace in that copy receives
each contribution, while retaining the containment sums.
To make this choice, represent each crossing subspace by two
coordinates: its direction identifies the subspace inside a copy,
and Fourier transformation of its translation determines the
contributions to the copies.

\begin{lem}[Crossing subspaces]\label{lem:crossing}
Every subspace $U\in\mathcal X_H$ has a unique pair $(B,x+B)$ such that
\begin{equation}\label{eq:lift}
 U=U(B,x+B):=(B\times\{0\})+
       \left\langle\begin{pmatrix}x\\1\end{pmatrix}\right\rangle,
 \qquad B\le\F_q^{n-1},\quad x+B\in\F_q^{n-1}/B.
\end{equation}
Here $B\times\{0\}=U\cap H$ and $\dim U=\dim B+1$.
For $U_i=U(B_i,x_i+B_i)$, $i=1,2$,
\[
 U_1\le U_2\quad\Longleftrightarrow\quad
 B_1\le B_2\ \text{ and }\ x_1-x_2\in B_2.
\]
\end{lem}

The condition $B_1\le B_2$ is exactly what we need to apply a smaller
zeta transform inside each frequency copy. Keep $B$ as the subspace coordinate and apply Fourier
transformation to $x+B$ to remove the second condition from the sum
while keeping $B_1\le B_2$.
For complex weights, use the additive character $\psi$ from
\cref{sec:fourier-prelim}. Each $U(B,x+B)$ contributes
$f_1(U(B,x+B))\psi(\lambda(x))$ at subspace $B$ in copy $\lambda$,
whenever $B\le\ker\lambda$. Summing these contributions gives
\begin{equation}\label{eq:overview-fourier}
 \widehat f_\lambda(B)=
 \sum_{x+B\in\F_q^{n-1}/B}
     f_1\bigl(U(B,x+B)\bigr)\psi(\lambda(x)),
 \qquad B\le\ker\lambda.
\end{equation}

\begin{figure}[htbp]

\centering
 \resizebox{\linewidth}{!}{%
\begingroup
\definecolor{fftteal}{RGB}{0,134,118}
\definecolor{fftamber}{RGB}{188,111,18}
\definecolor{fftviolet}{RGB}{134,73,178}
\begin{tikzpicture}[x=1cm,y=1cm,
  card/.style={draw=black!38, rounded corners=1.1pt, minimum width=2.19cm, minimum height=.87cm, inner sep=1.1pt, font=\Large, align=center},
  lead/.style={-{Stealth[length=1.55mm,width=1.10mm]},line width=.63pt,opacity=.69,shorten <=1pt,shorten >=0.6pt},
  thru/.style={line width=.67pt,opacity=.56},
  fromfft/.style={-{Stealth[length=1.45mm,width=1.0mm]},line width=.68pt,opacity=.78,shorten >=1.5pt},
]
\node[font=\Large\bfseries,text=black!64] at (-9.205,1.98) {$B=0$};
\draw[black!16,line width=.5pt] (-14.200,1.74) -- (-4.210,1.74);
\node[font=\Large\bfseries,text=black!64] at (-1.315,1.98) {$B=\langle a\rangle$};
\draw[black!16,line width=.5pt] (-3.680,1.74) -- (1.050,1.74);
\node[font=\Large\bfseries,text=black!64] at (3.945,1.98) {$B=\langle b\rangle$};
\draw[black!16,line width=.5pt] (1.580,1.74) -- (6.310,1.74);
\node[font=\Large\bfseries,text=black!64] at (9.205,1.98) {$B=\langle a+b\rangle$};
\draw[black!16,line width=.5pt] (6.840,1.74) -- (11.570,1.74);
\node[font=\Large\bfseries,text=black!64] at (13.150,1.98) {$B=H$};
\draw[black!16,line width=.5pt] (12.100,1.74) -- (14.200,1.74);
\node[card,fill=blue!3] (c) at (-13.150,1.04) {$\langle c\rangle$};
\node[card,fill=blue!3] (ac) at (-10.520,1.04) {$\langle a+c\rangle$};
\node[card,fill=blue!3] (bc) at (-7.890,1.04) {$\langle b+c\rangle$};
\node[card,fill=blue!3] (abc) at (-5.260,1.04) {$\langle a+b+c\rangle$};
\node[card,fill=blue!3] (a_c) at (-2.630,1.04) {$\langle a,c\rangle$};
\node[card,fill=blue!3] (a_bc) at (0.000,1.04) {$\langle a,b+c\rangle$};
\node[card,fill=blue!3] (b_c) at (2.630,1.04) {$\langle b,c\rangle$};
\node[card,fill=blue!3] (b_ac) at (5.260,1.04) {$\langle b,a+c\rangle$};
\node[card,fill=blue!3] (ab_c) at (7.890,1.04) {$\langle a+b,c\rangle$};
\node[card,fill=blue!3] (ab_ac) at (10.520,1.04) {$\langle a+b,a+c\rangle$};
\node[card,fill=blue!3] (V) at (13.150,1.04) {$V$};
\fill[blue!5] (-13.555,-2.03) rectangle (-2.045,-2.90);
\draw[draw=blue!78!black,line width=.9pt,rounded corners=1.5pt] (-13.555,-2.03) rectangle (-2.045,-2.90);
\fill[fftteal!5] (-1.720,-2.03) rectangle (2.820,-2.90);
\draw[draw=fftteal!78!black,line width=.9pt,rounded corners=1.5pt] (-1.720,-2.03) rectangle (2.820,-2.90);
\fill[fftamber!5] (3.480,-2.03) rectangle (8.020,-2.90);
\draw[draw=fftamber!78!black,line width=.9pt,rounded corners=1.5pt] (3.480,-2.03) rectangle (8.020,-2.90);
\fill[fftviolet!5] (8.680,-2.03) rectangle (13.220,-2.90);
\draw[draw=fftviolet!78!black,line width=.9pt,rounded corners=1.5pt] (8.680,-2.03) rectangle (13.220,-2.90);
\draw[black!62,densely dashed,line width=.9pt,rounded corners=2.5pt] (-13.93,-1.73) rectangle (13.59,-3.20);
\node[anchor=west,fill=white,inner xsep=3pt,inner ysep=1pt,font=\Large\bfseries] at (-13.78,-3.20) {FFT};
\node[card,fill=blue!6,draw=blue!58] (t1_0) at (-12.450,-4.25) {$0$};
\node[card,fill=blue!6,draw=blue!58] (t1_a) at (-10.125,-4.25) {$\langle a\rangle$};
\node[card,fill=blue!6,draw=blue!58] (t1_b) at (-7.800,-4.25) {$\langle b\rangle$};
\node[card,fill=blue!6,draw=blue!58] (t1_ab) at (-5.475,-4.25) {$\langle a+b\rangle$};
\node[card,fill=blue!6,draw=blue!58] (t1_H) at (-3.150,-4.25) {$H$};
\node[font=\Large\bfseries,text=blue!82!black] at (-7.800,-5.08) {$\operatorname{Gr}_2(2)$};
\node[card,fill=fftteal!6,draw=fftteal!58] (t2_0) at (-0.615,-4.25) {$0$};
\node[card,fill=fftteal!6,draw=fftteal!58] (t2_a) at (1.715,-4.25) {$\langle a\rangle$};
\node[font=\Large\bfseries,text=fftteal!82!black] at (0.550,-5.08) {$\operatorname{Gr}_2(1)$};
\node[card,fill=fftamber!6,draw=fftamber!58] (t3_0) at (4.585,-4.25) {$0$};
\node[card,fill=fftamber!6,draw=fftamber!58] (t3_b) at (6.915,-4.25) {$\langle b\rangle$};
\node[font=\Large\bfseries,text=fftamber!82!black] at (5.750,-5.08) {$\operatorname{Gr}_2(1)$};
\node[card,fill=fftviolet!6,draw=fftviolet!58] (t4_0) at (9.785,-4.25) {$0$};
\node[card,fill=fftviolet!6,draw=fftviolet!58] (t4_ab) at (12.115,-4.25) {$\langle a+b\rangle$};
\node[font=\Large\bfseries,text=fftviolet!82!black] at (10.950,-5.08) {$\operatorname{Gr}_2(1)$};
\begin{scope}[on background layer]
\draw[lead,draw=blue!88!black] ([xshift=-0.420cm]c.south) -- (-12.450,-2.03);
\draw[lead,draw=fftteal!88!black] ([xshift=-0.140cm]c.south) -- (-0.615,-2.03);
\draw[lead,draw=fftamber!88!black] ([xshift=0.140cm]c.south) -- (4.585,-2.03);
\draw[lead,draw=fftviolet!88!black] ([xshift=0.420cm]c.south) -- (9.785,-2.03);
\draw[lead,draw=blue!88!black] ([xshift=-0.420cm]ac.south) -- (-12.450,-2.03);
\draw[lead,draw=fftteal!88!black] ([xshift=-0.140cm]ac.south) -- (-0.615,-2.03);
\draw[lead,draw=fftamber!88!black,dashed,dash pattern=on 2pt off 1.6pt] ([xshift=0.140cm]ac.south) -- (4.585,-2.03);
\draw[lead,draw=fftviolet!88!black,dashed,dash pattern=on 2pt off 1.6pt] ([xshift=0.420cm]ac.south) -- (9.785,-2.03);
\draw[lead,draw=blue!88!black] ([xshift=-0.420cm]bc.south) -- (-12.450,-2.03);
\draw[lead,draw=fftteal!88!black,dashed,dash pattern=on 2pt off 1.6pt] ([xshift=-0.140cm]bc.south) -- (-0.615,-2.03);
\draw[lead,draw=fftamber!88!black] ([xshift=0.140cm]bc.south) -- (4.585,-2.03);
\draw[lead,draw=fftviolet!88!black,dashed,dash pattern=on 2pt off 1.6pt] ([xshift=0.420cm]bc.south) -- (9.785,-2.03);
\draw[lead,draw=blue!88!black] ([xshift=-0.420cm]abc.south) -- (-12.450,-2.03);
\draw[lead,draw=fftteal!88!black,dashed,dash pattern=on 2pt off 1.6pt] ([xshift=-0.140cm]abc.south) -- (-0.615,-2.03);
\draw[lead,draw=fftamber!88!black,dashed,dash pattern=on 2pt off 1.6pt] ([xshift=0.140cm]abc.south) -- (4.585,-2.03);
\draw[lead,draw=fftviolet!88!black] ([xshift=0.420cm]abc.south) -- (9.785,-2.03);
\draw[lead,draw=blue!88!black] ([xshift=-0.150cm]a_c.south) -- (-10.125,-2.03);
\draw[lead,draw=fftteal!88!black] ([xshift=0.150cm]a_c.south) -- (1.715,-2.03);
\draw[lead,draw=blue!88!black] ([xshift=-0.150cm]a_bc.south) -- (-10.125,-2.03);
\draw[lead,draw=fftteal!88!black,dashed,dash pattern=on 2pt off 1.6pt] ([xshift=0.150cm]a_bc.south) -- (1.715,-2.03);
\draw[lead,draw=blue!88!black] ([xshift=-0.150cm]b_c.south) -- (-7.800,-2.03);
\draw[lead,draw=fftamber!88!black] ([xshift=0.150cm]b_c.south) -- (6.915,-2.03);
\draw[lead,draw=blue!88!black] ([xshift=-0.150cm]b_ac.south) -- (-7.800,-2.03);
\draw[lead,draw=fftamber!88!black,dashed,dash pattern=on 2pt off 1.6pt] ([xshift=0.150cm]b_ac.south) -- (6.915,-2.03);
\draw[lead,draw=blue!88!black] ([xshift=-0.150cm]ab_c.south) -- (-5.475,-2.03);
\draw[lead,draw=fftviolet!88!black] ([xshift=0.150cm]ab_c.south) -- (12.115,-2.03);
\draw[lead,draw=blue!88!black] ([xshift=-0.150cm]ab_ac.south) -- (-5.475,-2.03);
\draw[lead,draw=fftviolet!88!black,dashed,dash pattern=on 2pt off 1.6pt] ([xshift=0.150cm]ab_ac.south) -- (12.115,-2.03);
\draw[lead,draw=blue!88!black] ([xshift=0.000cm]V.south) -- (-3.150,-2.03);
\end{scope}
\draw[thru,draw=blue!86!black] (-12.450,-2.03) -- (-12.450,-2.90);
\draw[fromfft,draw=blue!88!black] (-12.450,-2.90) -- (t1_0.north);
\draw[thru,draw=blue!86!black] (-10.125,-2.03) -- (-10.125,-2.90);
\draw[fromfft,draw=blue!88!black] (-10.125,-2.90) -- (t1_a.north);
\draw[thru,draw=blue!86!black] (-7.800,-2.03) -- (-7.800,-2.90);
\draw[fromfft,draw=blue!88!black] (-7.800,-2.90) -- (t1_b.north);
\draw[thru,draw=blue!86!black] (-5.475,-2.03) -- (-5.475,-2.90);
\draw[fromfft,draw=blue!88!black] (-5.475,-2.90) -- (t1_ab.north);
\draw[thru,draw=blue!86!black] (-3.150,-2.03) -- (-3.150,-2.90);
\draw[fromfft,draw=blue!88!black] (-3.150,-2.90) -- (t1_H.north);
\draw[thru,draw=fftteal!86!black] (-0.615,-2.03) -- (-0.615,-2.90);
\draw[fromfft,draw=fftteal!88!black] (-0.615,-2.90) -- (t2_0.north);
\draw[thru,draw=fftteal!86!black] (1.715,-2.03) -- (1.715,-2.90);
\draw[fromfft,draw=fftteal!88!black] (1.715,-2.90) -- (t2_a.north);
\draw[thru,draw=fftamber!86!black] (4.585,-2.03) -- (4.585,-2.90);
\draw[fromfft,draw=fftamber!88!black] (4.585,-2.90) -- (t3_0.north);
\draw[thru,draw=fftamber!86!black] (6.915,-2.03) -- (6.915,-2.90);
\draw[fromfft,draw=fftamber!88!black] (6.915,-2.90) -- (t3_b.north);
\draw[thru,draw=fftviolet!86!black] (9.785,-2.03) -- (9.785,-2.90);
\draw[fromfft,draw=fftviolet!88!black] (9.785,-2.90) -- (t4_0.north);
\draw[thru,draw=fftviolet!86!black] (12.115,-2.03) -- (12.115,-2.90);
\draw[fromfft,draw=fftviolet!88!black] (12.115,-2.90) -- (t4_ab.north);
\fill[blue!85!black] (-12.450,-2.03) circle[radius=.042cm];
\fill[blue!85!black] (-10.125,-2.03) circle[radius=.042cm];
\fill[blue!85!black] (-7.800,-2.03) circle[radius=.042cm];
\fill[blue!85!black] (-5.475,-2.03) circle[radius=.042cm];
\fill[blue!85!black] (-3.150,-2.03) circle[radius=.042cm];
\fill[fftteal!85!black] (-0.615,-2.03) circle[radius=.042cm];
\fill[fftteal!85!black] (1.715,-2.03) circle[radius=.042cm];
\fill[fftamber!85!black] (4.585,-2.03) circle[radius=.042cm];
\fill[fftamber!85!black] (6.915,-2.03) circle[radius=.042cm];
\fill[fftviolet!85!black] (9.785,-2.03) circle[radius=.042cm];
\fill[fftviolet!85!black] (12.115,-2.03) circle[radius=.042cm];
\node[font=\large\bfseries,text=blue!80!black,inner sep=0pt] at (-2.565,-2.460) {$\lambda_1$};
\node[font=\large\bfseries,text=fftteal!80!black,inner sep=0pt] at (2.300,-2.460) {$\lambda_2$};
\node[font=\large\bfseries,text=fftamber!80!black,inner sep=0pt] at (7.500,-2.460) {$\lambda_3$};
\node[font=\large\bfseries,text=fftviolet!80!black,inner sep=0pt] at (12.700,-2.460) {$\lambda_4$};
\end{tikzpicture}
\endgroup
 } 
\caption[Fourier coordinates for the crossing subspaces of $\F_2^3$]{%
Fourier decomposition in $\F_2^3=\langle a,b,c\rangle$, with
$H=\langle a,b\rangle$. The top row groups crossing subspaces
$U=B+\langle c+x\rangle$ by $B=U\cap H$, and Fourier transforms on $H/B$
combine their weights with signs $(-1)^{\lambda(x)}$, shown by solid
$(+1)$ and dashed $(-1)$ arrows. Regrouping the coefficients by frequency
$\lambda$ gives the bottom blocks, indexed by $B\le\ker\lambda$ with
their induced containment order: within each block, every subspace is
drawn to the left of those that strictly contain it.}
\label{fig:fourier-step}

\end{figure}
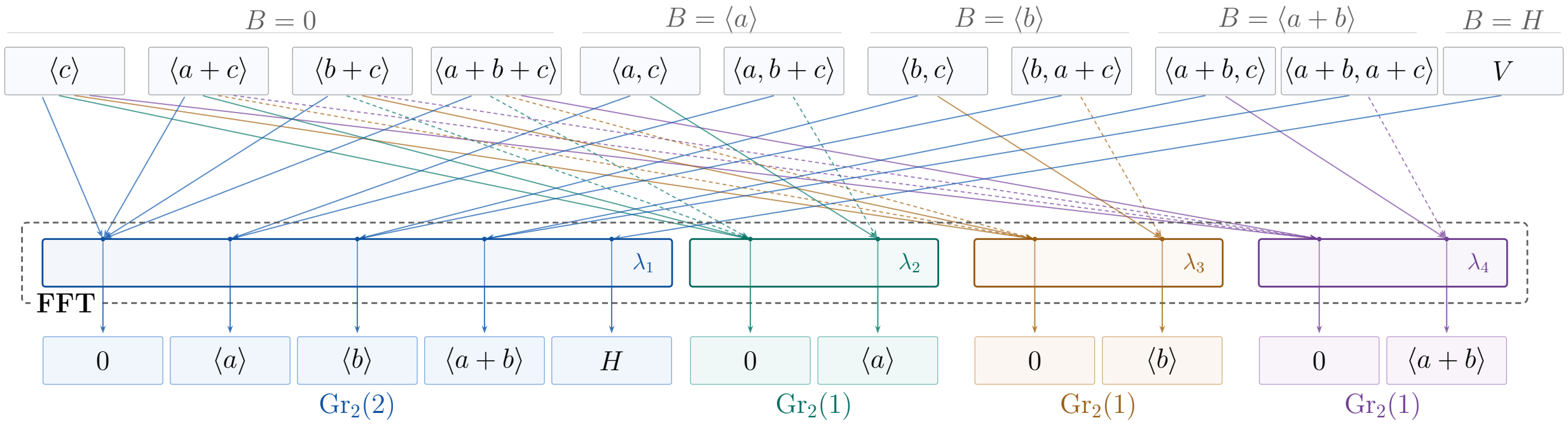

Figure~\ref{fig:fourier-step} illustrates this change of coordinates
for $\F_2^3$.

Fourier inversion recovers each table on $\F_q^{n-1}/B$, so all the
copies together retain the crossing weights. Let $F_B$ denote the
Fourier transform in \eqref{eq:overview-fourier} for fixed $B$.
If $P$ reorders
the combined output indices $(B,\lambda)$ as $(\lambda,B)$, then
\cref{lem:affine} gives
\begin{align*}
 \mathcal F&=P\left(\bigoplus_{B\le\F_q^{n-1}}F_B\right),\notag\\
 \AffineTransform&=\mathcal F^{-1}
 \operatorname{diag}\!\left(\ZetaTransform_{n-1},
 \underbrace{\ZetaTransform_{n-2},\ldots,\ZetaTransform_{n-2}}_{q^{n-1}-1\text{ copies}}\right)
 \mathcal F.
\end{align*}
These equations give a reversible map with the required containment
sums in each copy.
Algorithm~\ref{alg:sum} summarizes this recursion. A chosen basis
identifies each kernel with its coordinate space.

 \begin{algorithm}[htbp]
\caption{The full zeta transform: $\operatorname{Sum}(n,f)=\ZetaTransform_n f$}\label{alg:sum}
\KwIn{A table $f$ on $\Gr(n)$.}
\KwOut{The table $\ZetaTransform_n f$.}
\lIf{$n=0$}{\Return $f$}
Split $f$ into $f_0$ on the subspaces inside $H$ and $f_1$ on $\mathcal X_H$\;
$z_0\gets\operatorname{Sum}(n-1,f_0)$\;
$(\widehat f_\lambda)_{\lambda\in\F_q^{n-1}}\gets\mathcal F f_1$\;
\ForEach{$\lambda\in\F_q^{n-1}$}{
$\widehat u_\lambda\gets\operatorname{Sum}(\dim\ker\lambda,\widehat f_\lambda)$\;
}
$u\gets\mathcal F^{-1}\left((\widehat u_\lambda)_{\lambda\in\F_q^{n-1}}\right)$\;
\Return %
$(\ZetaTransform_n f)(W)=\begin{cases}
z_0(W),&W\le H,\\
z_0(W\cap H)+u(W),&W\in\mathcal X_H.
\end{cases}$\;
\end{algorithm}

\paragraph{Dimension cutoffs and running time.}

We now extend the recursion to $\ZetaTransform_{n,k}$, which sums over
subspaces of dimension at most $k$. The hyperplane partition still
works: inside $H$ the cutoff stays $k$, while
for a crossing subspace $U(B,x+B)$, the condition $\dim U\le k$
becomes $\dim B\le k-1$. Thus the recursive transforms are
\[
 \ZetaTransform_{n-1,k}f_0
 \quad\text{and}\quad
 \ZetaTransform_{\dim\ker\lambda,k-1}\widehat f_\lambda
       \quad(\lambda\in\F_q^{n-1}).
\]
To bound the total work, first observe that the copying has not
increased the number of entries. For each $B$, there are equally many
cosets $x+B$ and frequencies with $\lambda(B)=0$. The smaller
transforms therefore have $M_q(n,k)$ entries in total, as expressed
by \eqref{eq:count}, and the Fourier steps cost $O_q(nM_q(n,k))$.

We prove a total bound proportional to $nkM_q(n,k)$ by induction.
The saving comes from the calls with cutoff $k-1$, which contain
at least half the entries. For these entries, the factor $nk$
decreases by at least $nk-(n-1)(k-1)\ge n$, and for the others it does
not increase. This decrease covers the Fourier work after choosing
a sufficiently large constant. Thus the total cost is
$O_q(nkM_q(n,k))=O_q(M_q(n,k)\log M_q(n,k))$,
as detailed in \cref{sec:linear}.

\paragraph{Inversion and incidence.}
To compute $\ZetaTransform_{n,k}^{-1}$, recover $f$ from the supplied
output $(z_0,z_1)$ by solving the block equations in
\eqref{eq:overview-cut}, with the dimension cutoff, in order:
\begin{equation}\label{eq:inverse}
 f_0=\ZetaTransform_{n-1,k}^{-1}z_0,\qquad
 f_1=\AffineTransform^{-1}\bigl(z_1-Cf_0\bigr).
\end{equation}
The inverse of $\AffineTransform$ uses the same Fourier maps with
inverse smaller zeta transforms. To obtain sums over superspaces
as well, we transpose these linear computations: each weighted
contribution is sent in the reverse direction and contributions
at each input are added. This gives $\ZetaTransform_{n,k}^{\mathsf T}$
and its inverse with the same order of work.

For the incidence transform $\IncidenceTransform_n^{a,b}$, retain
only $a$-dimensional inputs and $b$-dimensional outputs in the same
hyperplane partition. Let $\AffineTransform_{a-1,b-1}$ sum over crossing
inputs with directions of dimension $a-1$ and crossing outputs with
directions of dimension $b-1$. For inside and crossing inputs $f_0,f_1$,
the output $z=(z_0,z_1)$ satisfies
\begin{align*}
 z_0&=\IncidenceTransform_{n-1}^{a,b}f_0,\\
 z_1\bigl(U(B,x+B)\bigr)
   &=(\IncidenceTransform_{n-1}^{a,b-1}f_0)(B)
     +(\AffineTransform_{a-1,b-1}f_1)(B,x+B).
\end{align*}
Fourier transformation replaces $\AffineTransform_{a-1,b-1}$
by $\IncidenceTransform_{\dim\ker\lambda}^{a-1,b-1}$ at each frequency.
The inside input $f_0$ is used in two calls, so \cref{app:incidence-work}
accounts separately for this repetition and proves the $O_q(S\log S)$
bound for $S=|\Gr(n,a)|+|\Gr(n,b)|$.
To extend the argument beyond complex weights, we must supply Fourier
character values that may not lie in $\mathcal R$.
Section~\ref{sec:arithmetic} represents them using a constant number
of elements of $\mathcal R$, so the same recursion runs at constant
overhead. Its outputs are the required sums in $\mathcal R$, and
its only divisions are by powers of $p$, as permitted in \cref{thm:linear}.

\subsubsection{Sums on Grassmann graphs}
To compute $Kf$ by two containment transforms, we will express each
weight $K(U,W)$ as a sum over subspaces of $U\cap W$.
Suppose $k\le n/2$. An upper transform of
$f$ on $\Gr(n,k)$ gives
$H(L)=\sum_{W\supseteq L,\,\dim W=k}f(W)$.
Summing over the $j$-spaces $L\le U$ counts each $W$ once for every
$j$-space in $U\cap W$:
\[
 \sum_{\substack{L\le U\\\dim L=j}}H(L)
   =\sum_{W\in\Gr(n,k)}|\Gr(\dim(U\cap W),j)| f(W).
\]
The matrix $(|\Gr(t,j)|)_{j,t=0}^k$ is triangular with diagonal one.
To apply $K$, choose $c_0,\ldots,c_k$ by back-substitution so that
$a_{k-t}=\sum_{j=0}^t|\Gr(t,j)| c_j$. Then
\[
 (Kf)(U)=\sum_{L\le U}c_{\dim L}H(L).
\]
Thus we apply $K$ by an upper transform, multiplication of each
$H(L)$ by $c_{\dim L}$, and a lower transform.
For $k>n/2$, taking annihilators preserves
distances and reduces to the preceding case.

 Over $\Q$ or $\C$, interpolate $P_j$ with $P_j(\theta_i)=\ind[i=j]$
 at the distinct eigenvalues of $D_1$. Then $P_j(D_1)f$ is the
$\theta_j$-component. Over fields with invertible $p$, symmetry makes
$K^{-1}$ an intersection kernel whenever $K$ is nonsingular. Its
$d+1$ coefficients follow from a linear system of that size.
Appendix~\ref{app:grassmann} proves these consequences.

\subsubsection{Counting subsets by exact span}
Given vectors $v_1,\ldots,v_\ell\in\F_q^n$, let $C(U)$ count the
subsets of indices whose vectors span $U$. Weight each subspace by the
number of input vectors spanning it---only the zero subspace and lines
can have nonzero weight. A zeta transform gives
$a(U)=|\{i:v_i\in U\}|$. Every subset using only these vectors has a
unique span contained in $U$, so
\[
 2^{a(U)}=\sum_{V\le U}C(V).
\]
One M\"obius transform therefore recovers all $C(U)$, with the running
time given in \cref{thm:span-count}.

\subsubsection{Graph coloring}
To decide which classes can share a color across a cut $(X,Y)$
of rank $r$, factor its adjacency matrix to give
labels $p_x,q_y\in\F_2^r$ with $xy\in E$ exactly when
$p_x^{\mathsf T}q_y=1$. Two color classes on opposite sides can share
a color exactly when the spans of their labels are orthogonal.
Call these spans the classes' profiles.

First consider one coloring of each child of a decomposition node,
with child vertex sets $X_1,X_2$. A parent coloring pairs some classes
from the left child with classes from the right child, leaving unpaired
classes separate. Each pair must have no edge between its
two classes, a condition determined by their profiles. Let $\pi_i$
restrict a child's neighborhood vectors to vertices outside
$X_1\cup X_2$. A paired class with child profiles $U,V$ has parent
profile $\pi_1(U)+\pi_2(V)$, while unpaired classes have profiles
$\pi_1(U)$ and $\pi_2(V)$, respectively.
Every parent color partition determines its child partitions and
this partial matching uniquely.

We now compute all possible parent histograms (counts of classes
with each profile) from the child histograms.
Write $L=N_2(r)$. Recording a multiplicity for every pair of profiles
would require up to $L^2$ counts. Instead, fix a left histogram and
process its classes one at a time. Each class either stays separate
or takes an available compatible right partner. We need only the
numbers of unused right classes and the histogram of parent classes
already formed. These are at most $2L$ counts, each at most the number
of vertices $N$. There are $N^{O(L)}$ states and left histograms, at most
$N$ steps, and at most $L+1$ choices per step, giving $N^{O(L)}$
 total work. After the last left class, the remaining right classes
stay separate. Choosing a partner with profile $V$ has weight equal
to the number of unused right classes of that profile, so the same
recurrence counts partitions.

 At the root, let $a_t$ count the partitions into $t$ nonempty
independent classes. Assigning distinct colors to these classes
gives the chromatic polynomial
\[
 P_G(z)=\sum_{t=0}^N a_t(z)_t,\qquad
 (z)_t=z(z-1)\cdots(z-t+1),\quad (z)_0=1.
\]
The chromatic number is the least $t$ with $a_t>0$.
Section~\ref{sec:coloring} proves the recurrence and its counting weights.

For the fixed-palette lower bound, arrange $k^2$ Boolean variables in
$k$ rows. Give an assignment $u\in\F_2^k$ to row $j$ the label
$(e_j,u)\in\F_2^{2k}$, and an assignment $v\in\F_2^k$ to column
$h$ the label $(v,e_h)$. Orthogonality is precisely the consistency
condition $u_h=v_j$. Selecting at least one assignment in each row
and column therefore forces one matrix of assignments, shared by
all clause copies. Odd cycles enforce these nonempty choices and
have exactly two extensions for each forced singleton selection.
Constant-size clause gadgets give sixteen extensions for every
satisfied clause. The same graph thus proves both the decision and
exact counting bounds, with cut-rank $O(k)$.

For unrestricted palettes, we represent integer variables by numbers
of colors. Use the $2^{d^2}$ subspaces
$\{(x,A^{\mathsf T}x):x\in\F_2^d\}$, one for each binary
$d\times d$ matrix $A$, as color types, and let the number of colors
of each type represent one variable. Orthogonality tests let a
constraint use colors of a specified type, with total cut-rank
$2d+O(1)$. A known ETH lower bound excludes $f(K)m^{o(K)}$
time for the source problem with $K$ counters and domain size
$m$~\cite[Lemma~3.1]{FGL+19}. Choosing
$d=\lceil\sqrt{\log_2 K}\rceil$ provides enough types and gives
$r=2\sqrt{\log_2 K}+O(1)$, yielding the coefficient $1/4$ in the exponent.
Keeping the explicit additive constant in this width bound gives the
sharper exclusion in \cref{xp:lower}.
The two reductions are detailed in
\cref{fix:implementation,xp:section}.

  The reductions give quantitative implications without assuming ETH.
For example, a $T(N,r)$-time algorithm for three-colorability would
decide a $k^2$-variable $3$-CNF formula with $m$ clauses in time
\[
 2^{O(k)}\poly(m)+T\bigl(2^{O(k)}\poly(m),O(k)\bigr).
\]
ETH is used only to exclude the resulting SAT running times.

\paragraph{Related Work}

\begingroup
The covering-product identity of Bj{\"o}rklund, Husfeldt, Kaski, and Koivisto~\cite{BHKK07}
gives span convolution from subspace containment
sums. The covering-pair algorithm of~\cite{KKW16}
uses one operation per
covering pair on semimodular lattices.

Srinivasan~\cite[Theorems~1.4 and~2.5]{Sri14} uses translation characters
to decompose the operator summing over immediate superspaces.
For $B\le\ker\lambda$, his unnormalized character projection of
$U(B,B)$ has coefficient $q^{\dim B}\psi(-\lambda(x))$ at $U(B,x+B)$.
Dividing by $q^{n-1}$ gives the inverse Fourier column in \eqref{eq:ifft}.
We bound the cost of containment transforms on arbitrary tables,
including cutoffs and two prescribed dimensions, accounting for
Fourier coordinate changes and recursive calls over rings with
invertible $p$.

For coloring, we use the subspace profiles of~\cite{BTV10,GHO13}.
Our transform answers compatibility queries on supplied profile tables, whereas
the coloring algorithm instead merges profile multiplicities.
Ganian, Hlin\textcaronaccent{e}n\'y, and Obdr\textcaronaccent{z}\'alek~\cite[Theorems~4.1 and~4.7]{GHO13}
enumerate matching multiplicities for pairs of profiles to compute
the chromatic number and polynomial. Our recurrence processes the left classes one profile at a time.

Bergougnoux, Korhonen, and Nederlof~\cite{BKN23} ask for tight rank-width
bounds for fixed-palette coloring and Chromatic Number. Their reduction
encodes assignments by subspaces and forces consistency through a quota
of common nonneighbors~\cite[Lemma~4.9, full version]{BKN23}. We retain
the assignment encoding and enforce consistency by testing agreement
between row and column assignments. Odd cycles and fixed-size clause
gadgets realize these choices in three-coloring with cut-rank $O(k)$
for $k^2$ SAT variables, and give a fixed number of extensions per
satisfying assignment.

 For unrestricted palettes, we use the monotone-counter constraints and
selector of Fomin, Golovach, Lokshtanov,
Saurabh, and Zehavi~\cite{FGL+19}. Our matrix register and type tests
store $K$ counters at cut-rank $2\sqrt{\log_2 K}+O(1)$, yielding the
matching coefficient $1/4$.
 \par
\endgroup
\section{Preliminaries and definitions}\label{sec:prelim}
We interpret $\Gr(n,j)=\varnothing$ when $j<0$ or $j>n$ and use
empty tables for negative cutoffs.
Choose a basis for each space and write vectors as columns. For subspaces,
$U+W=\{u+w:u\in U,w\in W\}$, and $U\oplus W$ means $U\cap W=\{0\}$.
For $B\le H=\F_q^d$, an affine flat $x+B$ is a coset of its direction $B$.
The cosets form the vector space $H/B$, with addition
$(x+B)+(y+B)=(x+y)+B$ and dimension $\dim H-\dim B$.
Two cosets agree exactly when their representatives differ by an element of $B$.
We identify $\lambda\in\F_q^d$ with the linear map
$\lambda(x)=\lambda^{\mathsf T}x$ and write
$\ker\lambda=\{x:\lambda(x)=0\}$.
For $B\le\F_q^d$, its \emph{annihilator} is
$B^\circ=\{\lambda\in\F_q^d:\lambda(B)=0\}$, of dimension $d-\dim B$.

A \emph{table} on $X$ is a function $f:X\to\mathcal R$, viewed as a vector in $\mathcal R^X$.
Matrix rows index outputs and columns index inputs. Ordering subspaces
by increasing dimension makes the containment matrix triangular with
diagonal one, so its inverse exists over every ring. Our fast algorithm
additionally requires that $p=\operatorname{char}\F_q$ be invertible.
We count ring additions, subtractions, and multiplications, and field
divisions for linear systems. Constants in $O_q(\cdot)$ depend only
on $q$. Appendices~\ref{sec:counts} and~\ref{app:implementation}
give the subspace counts, the construction of coordinates for the
recursive tables, and the bit costs.

\subsection{The finite Fourier transform}\label{sec:fourier-prelim}
We use the Fourier transform on each quotient $H/B$, with a nontrivial
additive character $\psi:\F_q\to\C$. For $q=p^m$, one choice is
$\psi(t)=\exp(2\pi i\operatorname{Tr}(t)/p)$, where
$\operatorname{Tr}(t)=t+t^p+\cdots+t^{p^{m-1}}\in\F_p$
and the exponential uses an integer representative.
For $B\le H=\F_q^d$ and $h_B:H/B\to\C$, Fourier transformation
and its inverse are
\begin{align}
 \widehat h_\lambda(B)&=\sum_{x+B\in H/B}
             h_B(x+B)\psi(\lambda(x)),
       &&\lambda\in B^\circ,\label{eq:fft}\\
 h_B(x+B)&=q^{-(d-\dim B)}\sum_{\lambda\in B^\circ}
             \widehat h_\lambda(B)\psi(-\lambda(x)).\label{eq:ifft}
\end{align}
The condition $\lambda(B)=0$ makes these formulas independent of coset
representatives, and both index sets have $q^{d-\dim B}$ entries.
Inversion follows from
\[
 \sum_{\lambda\in B^\circ}\psi(\lambda(z))
 =\begin{cases}q^{d-\dim B},&z\in B,\\0,&z\notin B.\end{cases}
\]
For the second case, $\lambda\mapsto\lambda(z)$ maps onto $\F_q$. Translating the resulting nontrivial character sum multiplies it by
a value different from one, so the sum vanishes.
To compute the transforms, identify $H/B$ with $\F_p^{m(d-\dim B)}$
and apply a length-$p$ transform along each coordinate, taking
$O_q((d+1)|H/B|)$ operations. Section~\ref{sec:arithmetic} implements
these formulas over the stated coefficient rings.

\section{A hyperplane cut and an ordinary Fourier transform}\label{sec:linear}
To prove \cref{thm:linear}, we extend the overview's decomposition to
subspaces of dimension at most $k$, working over $\C$ until \cref{sec:arithmetic}.

\subsection{The recursion with a dimension cutoff}
\begin{proof}[Proof of \cref{lem:crossing}]
With $H=\F_q^{n-1}\times\{0\}$, let $B\times\{0\}=U\cap H$.
Scale a vector of $U$ with nonzero last coordinate to
$\begin{psmallmatrix}x\\1\end{psmallmatrix}$. Subtracting multiples of
this vector leaves exactly $B\times\{0\}$, giving \eqref{eq:lift}.
Two choices of $x$ differ by an element of $B$, so $x+B$ is unique.
For $U_i=U(B_i,x_i+B_i)$, containment $U_1\le U_2$ holds exactly when
$B_1\le B_2$ and $\begin{psmallmatrix}x_1\\1\end{psmallmatrix}\in U_2$.
The latter condition is $x_1-x_2\in B_2$.
\end{proof}

We now compute $g=\ZetaTransform_{n,k}f$ recursively, using
$H=\F_q^{n-1}\times\{0\}$ and $U(B,x+B)$ from the overview.
Split the input into
$f_0(B)=f(B\times\{0\})$ and $(f_1)_B(x+B)=f(U(B,x+B))$,
and index the output parts $g_0,g_1$ in the same way.
To sum the crossing inputs, \cref{lem:crossing} requires sums over
contained cosets. For tables $h_B:\F_q^{n-1}/B\to\C$, define
\[
 (\AffineTransform h)_D(x+D)=\sum_{B\le D}\ \sum_{y+B\subseteq x+D}h_B(y+B).
\]
With cutoff $k$, inside subspaces have dimension at most $k$, and
crossing subspaces have directions of dimension at most $k-1$. Thus
\begin{equation}\label{eq:cut}
 g_0=\ZetaTransform_{n-1,k}f_0,\qquad
 (g_1)_D(x+D)=g_0(D)+(\AffineTransform f_1)_D(x+D).
\end{equation}
The value $g_0(D)$ comes from $\ZetaTransform_{n-1,k}f_0$.
It remains to compute $\AffineTransform f_1$, which we do by Fourier transformation.

\subsection{Why Fourier transformation removes translations}
To evaluate $\AffineTransform$, we apply \eqref{eq:fft} to each coset
table. The following identity leaves only the subspace containment
condition. We state it for $H=\F_q^d$ and tables $h_B:H/B\to\C$
with $\dim B\le\ell$.

\begin{lem}[Affine Fourier identity]\label{lem:affine}
For every direction $D$ of dimension at most $\ell$ and every
$\lambda\in D^\circ$,
\begin{equation}\label{eq:slice}
 \widehat{\AffineTransform h}_\lambda(D)=\sum_{B\le D}\widehat h_\lambda(B).
\end{equation}
\end{lem}
\begin{proof}
For each $B\le D$, every $B$-coset lies in exactly one $D$-coset, and
$\lambda$ is constant on that $D$-coset. Hence the total contribution
from direction $B$ is
\[
 \sum_{x+D}\sum_{y+B\subseteq x+D}h_B(y+B)\psi(\lambda(x))
 =\sum_{y+B}h_B(y+B)\psi(\lambda(y))=\widehat h_\lambda(B).
\]
Summing over $B$ proves the formula.
\end{proof}

For each frequency $\lambda$, the identity requires a zeta transform
on the subspaces of $\ker\lambda$. Let $\mathcal F$ apply \eqref{eq:fft}
for each $B$ and group the resulting entries by frequency. Order frequencies
with $0$ first, and order
each table $(\widehat h_\lambda(B))_{B\le\ker\lambda,\,\dim B\le\ell}$
by increasing dimension. The identity gives
\begin{equation}\label{eq:block}
 \AffineTransform=\mathcal F^{-1}
 \operatorname{diag}\!\left(\ZetaTransform_{d,\ell},
 \underbrace{\ZetaTransform_{d-1,\ell},\ldots,
 \ZetaTransform_{d-1,\ell}}_{q^d-1\text{ blocks}}\right)\mathcal F.
\end{equation}
A chosen basis identifies each kernel with its coordinate space,
giving one block per frequency even when kernels coincide. For $d=0$ there is only the first block.

Use $d=n-1$ and $\ell=k-1$ in \eqref{eq:block} to compute
$\AffineTransform f_1$, then add $g_0(D)$ as in \eqref{eq:cut}.
In each call, replace $k$ by $\min(k,n)$ and return the input if
$k=0$. All other calls decrease $n$.
For $\ZetaTransform_{n,k}^{-1}$, apply \eqref{eq:inverse} to $g=(g_0,g_1)$:
$Cf_0$ is the known value $g_0(D)$ at each crossing output $U(D,x+D)$.
Invert \eqref{eq:block} by using inverse smaller zeta transforms.

\subsection{The work follows the number of entries}
For each $B\le\F_q^{n-1}$, the $q^{n-1-\dim B}$ cosets correspond
to equally many frequencies with $\lambda(B)=0$. Thus the smaller
table lengths satisfy the
Goldman--Rota recurrence with a cutoff~\cite{Sri14}:
\begin{equation}\label{eq:count}
 M_q(n,k)=M_q(n-1,k)+M_q(n-1,k-1)
             +(q^{n-1}-1)M_q(n-2,k-1),\qquad n\ge2.
\end{equation}
By \cref{sec:fourier-prelim}, all local Fourier transforms together
take $O_q(nM_q(n,k))$ operations in a call with parameters $(n,k)$.

The case $k=0$ uses no arithmetic. For $k\ge1$, give each entry
a budget proportional to $nk$.
Only $\ZetaTransform_{n-1,k}f_0$ keeps cutoff $k$, whereas the transforms in
\eqref{eq:block} have cutoff $k-1$. At least half the entries belong
to these latter transforms, because
\[
 M_q(n,k)-2M_q(n-1,k)
 =(q^{n-1}-1)M_q(n-2,k-1)-|\Gr(n-1,k)|\ge0.
\]
For $1\le k<n$, use
$|\Gr(n-1,k)|=(q^{n-1}-1)|\Gr(n-2,k-1)|/(q^k-1)$. For $k=n$ the last count vanishes, and $n=1$ is direct.
For each such entry, the budget decreases by at least
$nk-(n-1)(k-1)=n+k-1\ge n$. All other entries
have nonincreasing budgets, leaving at least $nM_q(n,k)/2$ for the local
Fourier work. Induction gives the bound for the transform and its
inverse. Clipping each cutoff to its ambient dimension only reduces the budget.
For $1\le k\le n/2$, \eqref{eq:gaussian} gives
$\log_q M_q(n,k)\ge k(n-k)\ge nk/2$. For $k>n/2$, the middle-dimensional
term gives $\log_q M_q(n,k)\ge\lfloor n^2/4\rfloor=\Omega(nk)$.

To obtain sums over superspaces, apply the transposition
principle~\cite{BLS03} to the circuit for $\ZetaTransform_{n,k}$.
This computes $\ZetaTransform_{n,k}^{\mathsf T}$ with the same asymptotic cost.
Attach a zero accumulator $a_v$ to each circuit wire $v$. Add each
new input to the accumulator of its corresponding output wire, so
shared output wires receive the sum of these inputs. Traverse gates backwards:
\[
 z=x+y:\quad a_x\gets a_x+a_z,\ a_y\gets a_y+a_z,
 \qquad z=cx:\quad a_x\gets a_x+ca_z.
\]
The accumulators on the original input wires give the transpose.
Applying the same procedure to the inverse circuit computes
$(\ZetaTransform_{n,k}^{\mathsf T})^{-1}$, preserving the cutoff.

\subsection{Coefficient rings}\label{sec:arithmetic}
To run the recursion over a commutative ring $\mathcal R$ with
invertible $p$, we must supply the Fourier character values, which
need not lie in $\mathcal R$. For this, adjoin a formal variable
satisfying $1+t+\cdots+t^{p-1}=0$. Thus we work in
\[
 \mathcal R[t]/(1+t+\cdots+t^{p-1}),\qquad \zeta=t\bmod(1+t+\cdots+t^{p-1}),
 \qquad \psi(x)=\zeta^{\operatorname{Tr}_{\F_q/\F_p}(x)}.
\]
Each element is a polynomial reduced by this monic relation, hence
has a unique expansion in $1,t,\ldots,t^{p-2}$. Thus $\mathcal R$ embeds as
the constants and each extension operation costs
$O_q(1)$ operations in $\mathcal R$. Also $\zeta^p=1$ and
$\sum_{j=0}^{p-1}\zeta^{aj}=0$ for $a\not\equiv0\pmod p$.
These identities prove Fourier orthogonality even when $\mathcal R$
has zero divisors, and all divisions are by powers of $p$.
Since the outputs equal the original containment sums, all nonconstant
coefficients vanish, so extracting the constant coefficient gives
the answer in $\mathcal R$. This proves the coefficient statements
without assuming roots of unity in $\mathcal R$ or
dividing by subspace counts.

\section{Incidence between two dimensions}\label{sec:incidence}
Computing $\IncidenceTransform_n^{a,b}$ via the truncated zeta transform
$\ZetaTransform_{n,b}$ would require $M_q(n,b)$ entries, which can be
far more than $S:=|\Gr(n,a)|+|\Gr(n,b)|$.
To obtain the bound in \cref{thm:incidence}, which depends only on $S$, we retain only the two
required dimensions in the recursion. Let $f:\Gr(n,a)\to\C$ and
$g=\IncidenceTransform_n^{a,b}f$, so $g$ is indexed by $\Gr(n,b)$.
For $0<a<b<n$, cut at $H=\F_q^{n-1}\times\{0\}$.
Use $f_0,f_1,g_0,g_1$ as in \eqref{eq:cut}: the first parts are indexed
by subspaces in $H$, and the second by pairs $(B,x+B)$ representing $U(B,x+B)$.
The dimensions in $f_0,g_0$ are $a,b$, and the directions in $f_1,g_1$
have dimensions $a-1,b-1$. Consequently
\begin{align}
 g_0(D)&=(\IncidenceTransform_{n-1}^{a,b}f_0)(D),\notag\\
 (g_1)_D(x+D)&=(\IncidenceTransform_{n-1}^{a,b-1}f_0)(D)
       +\sum_{\substack{y+B\subseteq x+D\\\dim B=a-1}}(f_1)_B(y+B).
       \label{eq:rank-cut}
\end{align}
For $g_0$, only inputs inside the hyperplane can contribute. For a
crossing output with direction $D$, the inside inputs must lie in
$D$, explaining the second inside call with output dimension $b-1$.
The affine term separates by frequency as in \cref{lem:affine}, now
into incidence between dimensions $a-1,b-1$ in $\ker\lambda$.

Thus the algorithm Fourier-transforms each crossing input table
$(f_1)_B$, recursively applies
$\IncidenceTransform_{\dim\ker\lambda}^{a-1,b-1}$ at each frequency,
and uses \eqref{eq:ifft}, dividing by $q^{n-b}$, to recover
the affine sum. Add the second inside result as in \eqref{eq:rank-cut}.
For the boundary cases, return $f$ if $a=b$, copy $f(0)$
to every output if $a=0$, and return $\sum_Uf(U)$ if $b=n$.

\paragraph{Paying for the extra call.}
Equation~\eqref{eq:rank-cut} applies both $\IncidenceTransform_{n-1}^{a,b}$
and $\IncidenceTransform_{n-1}^{a,b-1}$ to $f_0$, so we must pay for
using the same input twice.
Let $T(n,a,b)$ be the arithmetic cost of $\IncidenceTransform_n^{a,b}$.
Input and output Fourier tables together
have at most $S$ entries, so
\begin{align}\label{eq:rank-work}
 T(n,a,b)\le{}&T(n-1,a,b)+T(n-1,a,b-1)+T(n-1,a-1,b-1)\notag\\
 &+(q^{n-1}-1)T(n-2,a-1,b-1)+O_q(nS).
\end{align}
The boundary cases cost $O(S)$. In every other call, all frequency
blocks have valid input and output dimensions, so listing frequencies is also
charged to their entries.

We pay for the extra call with a budget proportional to
  $n\min(a,n-b)$ per entry. The input dimension $a$ and output
codimension $n-b$ both decrease in the many nonzero-frequency calls, and
this decrease pays for the repeated inside call. The resulting bound
is $O_q(n\min(a,n-b)S)=O_q(S\log S)$ for the interior cases.
Appendix~\ref{app:incidence-work} gives the induction.
The ring extension in \cref{sec:arithmetic} and circuit transposition
then give the full statement of \cref{thm:incidence}.
\section{Convolution, exact spans, and intersection kernels}\label{sec:applications}
\subsection{Subspace convolution}
For sums over pairs with $U\cap V=W$, use the upper transform
$\ZetaTransform_n^{\mathsf T}$ in place of $\ZetaTransform_n$ in the
span-convolution argument from the introduction. This takes
$O_q(N_q(n)\log N_q(n))$ arithmetic operations.
For direct sums $U\oplus V=W$, apply span convolution with inputs
$f(U)z^{\dim U}$ and $g(V)z^{\dim V}$, and extract the coefficient
of $z^{\dim W}$.
Among pairs with $U+V=W$, the identity
$\dim U+\dim V=\dim W+\dim(U\cap V)$ shows that this coefficient
keeps exactly the pairs with $U\cap V=\{0\}$.
Coefficientwise transforms and schoolbook polynomial multiplication
give $O_q(n^3N_q(n))$ arithmetic operations. This is
the ranked-transform method of~\cite{BHKK07}, applied to subspaces.

\subsection{Counting subsets by their exact span}
M\"obius inversion also counts subsets with each possible span.

\begin{thm}[Counting exact spans]\label{thm:span-count}
Given $v_1,\ldots,v_\ell\in\F_q^n$, all numbers
\[
 C(U)=\#\{A\subseteq[\ell]:\operatorname{span}\{v_i:i\in A\}=U\},
 \qquad U\in\Gr(n),
\]
can be computed in $O_q(\ell+N_q(n)\log N_q(n))$ arithmetic
operations over $\Q$. The vectors may repeat and may be zero.
\end{thm}
\begin{proof}
For each subspace $B$, let $t(B)$ count the indices $i$ with
$\langle v_i\rangle=B$. This table is supported on the zero subspace
and the lines. Its zeta transform counts the input vectors in each
subspace: $a(U)=(\ZetaTransform_n t)(U)=|\{i:v_i\in U\}|$.
There are $2^{a(U)}$ subsets of these indices. Partitioning them by
their span gives
\begin{equation}\label{eq:exact-span}
 2^{a(U)}=\sum_{V\le U}C(V),\qquad
 C=\ZetaTransform_n^{-1}\bigl(U\mapsto2^{a(U)}\bigr).
\end{equation}
Precompute $2^0,\ldots,2^\ell$ using $\ell$ doublings and index this
list by $a(U)$.
Building $t$ takes $O(\ell)$ coefficient additions once the vectors are
indexed by their spans. The two transforms then give the bound.
Repeated vectors are distinct indexed choices, and
$C(\{0\})=2^{t(\{0\})}$ includes the empty subset.
\end{proof}

Normalizing and indexing the vectors takes $\ell\poly(n)$ operations over $\F_q$
for fixed $q$, in addition to $N_q(n)\poly(n)$ work for the transform
indices. Since the inversion input has at most $\ell+1$ bits per entry,
\cref{cor:integer-lifting} gives $(\ell+N_q(n))\poly(n,\ell)$ total bit
operations. For a uniformly chosen subset of the input indices,
$C(U)/2^\ell$ is the probability that its span is exactly $U$.

\subsection{Intersection kernels}
To compute $(Kf)(U)$, we express the weight of $f(W)$ as a sum over
subspaces of $U\cap W$, using the notation and coefficient assumption
of \cref{thm:kernels}.

Intersection kernels are exactly the matrices satisfying $K(gU,gW)=K(U,W)$ for
every invertible linear map $g\in\GL(n,q)$: extending bases sends
any pair to any other pair with the same intersection dimension.
To verify the distance formula, each edge changes the intersection
dimension with $W$ by at most one, and replacing the basis vectors of
$U$ outside $U\cap W$ by those of $W$ attains the bound.

\paragraph{A double-counting identity.}
 The cases $k=0,n$ have one entry and take $O(1)$ operations.
 Suppose $1\le k\le n/2$ and extend $f$ by zero to dimensions below $k$.
The upper transform gives
$H(L)=(\ZetaTransform_{n,k}^{\mathsf T}f)(L)=\sum_{W\in V:\,L\le W}f(W)$
for every $L$ with $\dim L\le k$.
For $U\in V$ and $0\le j\le k$, summing these values over the
$j$-dimensional subspaces of $U$ gives
\begin{equation}\label{eq:moments}
 m_j(U)=\sum_{\substack{L\le U\\\dim L=j}}H(L)
       =\sum_{W\in V}|\Gr(\dim(U\cap W),j)|f(W).
\end{equation}
 To assign weight $a_{k-t}$ to intersection dimension $t$, choose
$c_0,\ldots,c_k$ by
\begin{equation}\label{eq:kernel-coefficients}
 a_{k-t}=\sum_{j=0}^t|\Gr(t,j)|c_j,\qquad
 c_t=a_{k-t}-\sum_{j<t}|\Gr(t,j)|c_j.
\end{equation}
 Back-substitution takes $O(k^2)$ operations. Now weight each upper-transform value $H(L)$ by
$c_{\dim L}$ and apply one lower transform. At $U\in V$, its output is
\begin{equation}\label{eq:two-transforms}
 \sum_{L\le U}c_{\dim L}H(L)
 =\sum_{j=0}^k c_jm_j(U)=(Kf)(U).
\end{equation}
The subspace counts in \eqref{eq:kernel-coefficients} can be generated
by \eqref{eq:pascal} without division, so this works over every stated ring.
Since $M_q(n,k)=\Theta_q(N)$ and $\log N=\Theta_q(nk)$,
the two transforms and the $O(k^2)$ additional work take
$O_q(N\log N)$ operations.

For $k>n/2$, annihilators preserve distance, since
$\dim(U^\circ\cap W^\circ)=n-2k+\dim(U\cap W)$.
This proves \cref{thm:kernels} for all $k$.

 A dimension cutoff preserves intervals between retained subspaces,
so the earlier covering-pair transforms~\cite{KKW16} still use one addition
per covering pair. For $k\le n/2$, there are
$\sum_{j=1}^k|\Gr(n,j)|\qi j=\Theta_q(q^kN)$ such pairs:
successive terms grow by at least $q^2$. Using these transforms in
\eqref{eq:two-transforms}, and annihilators for $k>n/2$, gives
$O_q(q^dN)$ operations.

 \section{Subspace profiles and graph coloring}\label{sec:coloring}
To combine colorings across a cut, we record the span of the neighborhood
vectors of each color class, following~%
\cite{BTV10,GHO13}%
.
These subspace profiles let us test which classes can share a color.

\begin{thm}[Coloring on a supplied rank-decomposition]
\label{thm:general-coloring}\label{thm:chromatic}
Given an $N$-vertex graph and a rank-decomposition of width
at most $r$, its chromatic number, an optimal coloring, and its
chromatic polynomial can be computed in
\[
 N^{O(N_2(r))}=N^{2^{r^2/4+O(1)}}
\]
bit operations.
\end{thm}

\paragraph{Combining one pair of colorings.}
Suppress degree-two vertices of the decomposition and subdivide an
edge to root it as a binary tree. For the vertex set $X$ below a node,
the neighborhood vectors into $V(G)\setminus X$ span
\[
 W_X=\operatorname{col}(A_G[X,V(G)\setminus X]^{\mathsf T})
       \le\F_2^{V(G)\setminus X},\qquad \dim W_X\le r.
\]
A \emph{color partition} is a partition into nonempty independent
sets, with no names assigned to the classes. For a color partition of
$G[X]$, the profile of a class $C$ is
\[
 U_X(C)=\operatorname{span}\{A_G[x,V(G)\setminus X]^{\mathsf T}:x\in C\}
       \le W_X.
\]
Consider one color partition on each child set $X_1,X_2$, and put
$X=X_1\mathbin{\dot\cup}X_2$, $Y=V(G)\setminus X$, and $W_i=W_{X_i}$.
We must decide which classes can share a color and find their parent
profiles.

Restriction to $Y$ gives linear maps $\pi_i:W_i\to W_X$.
To test compatibility, define a bilinear form
$\beta:W_1\times W_2\to\F_2$. For representations
\[
 u=\sum_{x\in X_1}a_xA_G[x,X_2\cup Y]^{\mathsf T},\qquad
 v=\sum_{y\in X_2}b_yA_G[y,X_1\cup Y]^{\mathsf T},
\]
set $\beta(u,v)=\sum_{x\in X_1,\,y\in X_2}a_xb_yA_G[x,y]$.
Changing either representation adds a combination vanishing on the
opposite child, so $\beta$ is well-defined.

\begin{lem}[Compatibility and the parent profile]\label{lem:profile-merge}
Classes with profiles $U\le W_1$ and $V\le W_2$ can share a color
exactly when $\beta(U,V)=\{0\}$. Their union then has profile
$\pi_1(U)+\pi_2(V)$. The separate classes have profiles
$\pi_1(U)$ and $\pi_2(V)$, respectively.
\end{lem}
\begin{proof}
On the neighborhood vectors of vertices $x\in X_1$ and $y\in X_2$,
the value of $\beta$ is $A_G[x,y]$. Bilinearity makes its vanishing
on $U\times V$ equivalent to the absence of all edges between the
classes. Restricting their vectors to $Y$ gives the stated parent profiles.
\end{proof}

Classes sharing a color form a compatible partial matching: each
class has at most one partner. Conversely, such a matching gives a
parent partition. Restricting it to $X_1,X_2$ recovers both child
partitions and their matching, so the correspondence is a bijection.

\paragraph{Grouping partial colorings.}
Compatibility and parent profiles depend only on the child profiles.
We therefore group color partitions by the histogram
$h(U)=|\{C:U_X(C)=U\}|$, and store the number of color partitions realizing each histogram.
Nonempty classes with zero profile contribute to $h(\{0\})$, so $\sum_Uh(U)$ is the number of classes.

Fix a left histogram $h_1$. Rather than choose a multiplicity for
every pair of profiles, as in~\cite{GHO13}, process its classes one at a time, in a fixed
order of profiles. Retain two histograms: $b(V)$ counts right classes
of profile $V\le W_2$ still available, and $H(T)$ counts parent
classes of profile $T\le W_X$ already formed. For every right
histogram $h_2$, initialize $(b,H)=(h_2,0)$ with the product of the
two child counts.

 Write $\delta_T$ for the histogram counting one class of profile $T$.
  When the next left class has profile $U$, either leave it separate or
pair it with an available compatible right class:
\begin{align}
 (b,H)&\longmapsto(b,H+\delta_{\pi_1(U)}),
       &&\text{with weight }1,\label{eq:hist-row}\\
 (b,H)&\longmapsto(b-\delta_V,H+\delta_{\pi_1(U)+\pi_2(V)}),
       &&\text{with weight }b(V),\label{eq:hist-weight}
\end{align}
  where the second update requires $b(V)>0$ and $\beta(U,V)=\{0\}$.
Multiply each state's count by the transition weight and add it to
the count of the resulting state.
After all left classes have been processed, add each count to the parent
histogram $H+\sum_{V\le W_2}b(V)\delta_{\pi_2(V)}$.
  Sum these contributions over all left histograms $h_1$.

  To justify the weights, fix two concrete child partitions and order
the left classes of each profile by their least vertex. The next
left class is then fixed: it has one unmatched choice and exactly
$b(V)$ choices of a right partner with profile $V$. Each such choice
has the same new state, and later choices depend only on that state
and the remaining left profiles. Thus counts can be combined even
when they come from different right histograms. Every partial
matching has a unique sequence of these choices. The bijection above
therefore counts every parent partition once.

\paragraph{Initialization and output.}
At a leaf $\{v\}$, there is one partition, with one class of profile
$\operatorname{span}\{A_G[v,V(G)\setminus\{v\}]^{\mathsf T}\}$.
At the root, $W_{V(G)}=\{0\}$. The entry for $h(\{0\})=t$ counts
partitions into $t$ nonempty independent classes. Denote it by $a_t$.
Assigning distinct colors to the $t$ classes of each partition
gives the chromatic polynomial
\begin{equation}\label{eq:chromatic-falling}
 P_G(z)=\sum_{t=0}^N a_t(z)_t.
\end{equation}
Convert to ordinary coefficients by successively multiplying by $z-t$.
The chromatic number is the least $t$ with $a_t>0$.
Store one predecessor and pairing choice per positive state, and one
contributing left histogram and terminal state per parent histogram.
Backtracking recovers a right histogram and the pairing choices.
After recursively recovering the child partitions, follow these
choices using any available classes of the recorded profiles.
Lemma~\ref{lem:profile-merge} makes each pairing valid.

\begin{proof}[Running time in \cref{thm:general-coloring}]
Put $L=N_2(r)$. Each profile space has at most $L$ subspaces, and
each histogram entry is at most $N$. There are at most $(N+1)^L$
left histograms. For each one, the merge has at most $N$ steps,
$(N+1)^{2L}$ states per step, and $L+1$ transitions per state.
Over all nodes this gives $(N+1)^{3L}\poly(N,L)$ bit operations.
Gaussian elimination, subspace enumeration, and precomputing
restrictions, bilinear forms, compatibility tests, and profile sums
cost $\poly(N,L)$.
Pairs of partitions and partial matchings number $N^{O(N)}$, so all
stored counts have $O(N\log N)$ bits, and transition weights are at most
$N$. Conversion of \eqref{eq:chromatic-falling} uses the same bit
bound, so exact arithmetic costs a polynomial factor.

Since $\log L=O(L)$, the bound is $N^{O(L)}$, where
$L=2^{r^2/4+O(1)}$ by \cref{sec:counts}.
\end{proof}

\paragraph{Linear layouts.}\label{lin:section}
\begin{cor}\label{thm:linear-coloring}
On a supplied linear layout of width at most $r$, the three tasks
in \cref{thm:general-coloring} take
$(N+1)^{N_2(r)}2^{O(r^2)}N^{O(1)}$ bit operations.
\end{cor}
\begin{proof}
On a linear layout, adding one vertex lets us update one histogram
at a time. At a prefix $X$, let $x$ be the next vertex and
$R=V(G)\setminus(X\cup\{x\})$. On $W_X$, let $\ell$ evaluate the
$x$-coordinate, let $\pi$ restrict to $R$, and put
$b=A_G[x,R]^{\mathsf T}$. Deleting the $x$-coordinate sends a
histogram $h$ to $(\pi_*h)(V)=\sum_{U:\pi(U)=V}h(U)$.
The next vertex starts a new class or joins an existing one:
\begin{align}
 h&\longmapsto\pi_*h+\delta_{\operatorname{span}(b)},\label{lin:fresh}\\
 h&\longmapsto\pi_*(h-\delta_U)+\delta_{\pi(U)+\operatorname{span}(b)}
 \quad\text{if }h(U)>0\text{ and }\ell(U)=\{0\}.\label{lin:reuse}
\end{align}
Starting a new class has one choice, while joining a class of profile $U$
has $h(U)$ choices.
Starting with the empty histogram of count one, multiply by these
weights and add counts at equal histograms. Removing the last vertex
reverses exactly one choice, proving the counting invariant.
With $L=N_2(r)$, there are at most $(N+1)^L$ histograms,
$L+1$ choices per histogram, and $\poly(N,L)$ bit operations per choice.
\end{proof}
\paragraph{Compatibility queries on supplied profile tables.}
Labeled colorings across a binary rank-$r$ cut combine exactly when
corresponding profiles in $\F_2^r$ are orthogonal. Unused colors have
profile $\{0\}$.

\begin{prop}[Fast coloring-completion queries]\label{prop:color-compatibility}
Let $c\ge1$ and $S=N_2(r)^c$. Given vectors $f,g$ indexed by
$c$-tuples of subspaces of $\F_2^r$, with entries in a commutative ring
in which $2$ is invertible, all sums
\[
 B_g(U_1,\ldots,U_c)
 =\sum_{\substack{V_i\le U_i^\perp\\1\le i\le c}}g(V_1,\ldots,V_c)
\]
and $\sum_U f(U)B_g(U)$ take $O(S\log S)$ arithmetic operations.
Here $U^\perp$ uses the usual dot product.
\end{prop}
\begin{proof}
Apply $\ZetaTransform_r$ in each argument of $g$, holding the other
arguments fixed:
\begin{equation}\label{eq:color-compatibility}
 B_g(U_1,\ldots,U_c)
 =(\ZetaTransform_r^{\otimes c}g)(U_1^\perp,\ldots,U_c^\perp).
\end{equation}
These $cN_2(r)^{c-1}$ calls cost $O(cS\log N_2(r))=O(S\log S)$, and
the inner product costs $O(S)$.
\end{proof}
When $f$ and $g$ count labeled $c$-colorings of the two sides by their
ordered profile tuples, the final sum counts labeled $c$-colorings of $G$. With $H$-bit integer entries, the bit
cost over $\Q$ is $S\poly(c,r,H)$ by \cref{app:implementation}.
\section{Experiments}\label{sec:experiments}
Our algorithms give substantial speedups in all three applications
(Table~\ref{tab:exp-summary}).
The implementations, measurements, and scripts that generate the tables
and figure are available in the companion repository~\cite{PW26code}.

\begin{table}[H]

\centering\small
\begin{tabular}{@{}lccc@{}}
\toprule
Operation & Size & Speedup & \shortstack{Operation count\\ratio}\\
\midrule
Zeta transform $\ZetaTransform_n$ & $N_q(n):\ 1.2\cdot10^6$--$6.7\cdot10^8$ & $1.5$--$18$ & $0.45$--$4.1$\\
Incidence $\IncidenceTransform_n^{a,b}$ & $S:\ 1.8\cdot10^6$--$8.2\cdot10^8$ & $4.5$--$190$ & $4.3$--$170$\\
Grassmann kernel $Kf$ & $N:\ 2.6\cdot10^6$--$1.6\cdot10^8$ & $4.8$--$22$ & $0.81$--$4.6$\\
\bottomrule
\end{tabular}
\caption{Measured running time and operation count of the comparison algorithm
 divided by ours, for tests with at least $10^6$ entries and measured running
 times for both algorithms. Values above one favor our algorithms.}
\label{tab:exp-summary}

\end{table}

We compare the zeta transform $\ZetaTransform_n$ with the covering-pair algorithm~%
\cite{BHK+16,KKW16}%
,
and the incidence transform $\IncidenceTransform_n^{a,b}$ with the faster of
the covering-pair algorithm and Poisson summation~%
\cite[Theorem~10.1.16]{MP13}%
, using their measured running times.
For operation counts, we compare our incidence transform with direct
summation, the covering-pair algorithm, and Poisson summation, using the
smallest of their counts. For Grassmann kernels, we tested
$K(U,W)=2^{-\operatorname{dist}(U,W)}$ against the distance-matrix
recurrence~%
\cite[Section~5.3]{MOR03}%
, which computes $D_0f,\ldots,D_df$
before forming their weighted sum and uses incidence sums for each
adjacency multiplication.

The speedups often exceed the operation count ratios, and our algorithms
can have shorter measured running times despite using more operations.
This may partly reflect how the implementations read memory: our algorithm
stores corresponding
entries of the frequency subproblems next to one another, so many
additions and FFT steps scan contiguous rows, while the covering-pair
algorithm fetches a subspace value through a stored index for each addition.
For Grassmann kernels, Table~\ref{tab:exp-kernels} also reports measured
running times with the covering-pair algorithm in place of our fast
transforms in \eqref{eq:two-transforms}, showing how much of the speedup
comes from this identity and how much from our fast transforms.

\begin{figure}[tbp]

\centering
\includegraphics[width=\textwidth]{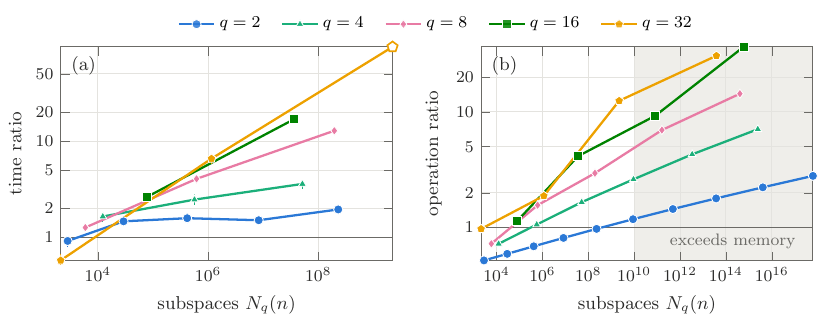}
\caption{Measured running time and operation count of the covering-pair algorithm
 divided by those of our zeta transform $\ZetaTransform_n$. Higher ratios favor
 our algorithm. Shading marks instances on which the covering-pair algorithm
 exceeds the available memory. For the yellow hollow marker, we estimated
 its running time by processing a sample of covering pairs and scaling to all pairs.}
\label{fig:exp-fields}

\end{figure}

Figure~\ref{fig:exp-fields} shows how the speedup for the zeta transform
$\ZetaTransform_n$ generally grows with table size and field size, though
the covering-pair algorithm can have shorter measured running times on
small tables. Poisson summation also has a shorter measured running time
on the point--hyperplane incidence test. Appendix~\ref{app:experiments}
gives the measurements and implementation details.

\label{page:main-end}
\clearpage
\appendix
\section{Counting subspaces}\label{sec:counts}
To count $r$-dimensional subspaces, with $0\le r\le n$, we count
their ordered bases: there are $\prod_{i=0}^{r-1}(q^n-q^i)$ ordered
linearly independent $r$-tuples in $\F_q^n$. Dividing by the number of bases
of each $r$-space gives
\begin{equation}\label{eq:gaussian}
 |\Gr(n,r)|=\prod_{i=0}^{r-1}\frac{q^n-q^i}{q^r-q^i}
 =q^{r(n-r)}\prod_{i=0}^{r-1}\frac{1-q^{i-n}}{1-q^{i-r}}
 =\Theta_q(q^{r(n-r)}).
\end{equation}
The estimate is uniform in $n,r$: the last product is between $1$
and $\prod_{i\ge1}(1-q^{-i})^{-1}<\infty$. Annihilators give
$|\Gr(n,r)|=|\Gr(n,n-r)|$. We also write
$\qi r=|\Gr(r,1)|=(q^r-1)/(q-1)$ for the number of lines in an
$r$-space. These integers can be computed without division, starting from
$|\Gr(0,0)|=1$, by
\begin{equation}\label{eq:pascal}
 |\Gr(n,r)|=|\Gr(n-1,r)|+q^{n-r}|\Gr(n-1,r-1)|.
\end{equation}
Indeed, the first term counts spaces in a fixed hyperplane, and each
$(r-1)$-space in that hyperplane has $q^{n-r}$ extensions outside it.

Completing the square in $r(n-r)$ gives
$N_q(n)=\Theta_q(q^{\lfloor n^2/4\rfloor})$. For dimensions at most $n/2$,
\begin{equation}\label{eq:truncated-size}
 M_q(n,k)=\Theta_q\!\left(|\Gr(n,k)|\right)\qquad(0\le k\le n/2),
\end{equation}
because the exponent for dimension $k-j$ is smaller than for dimension $k$ by
$j(n-2k)+j^2\ge j^2$, and $\sum_{j\ge0}q^{-j^2}$ is bounded.
The covering-pair algorithm processes pairs $U<W$ whose dimensions
differ by one. An $r$-dimensional subspace has
$\qi{n-r}$ immediate superspaces, one for each line in its quotient.
Their total number is
\begin{equation}\label{eq:cover-pairs}
 \sum_{r=0}^{n-1}|\Gr(n,r)|\qi{n-r}
 =\Theta_q\!\left(q^{\lfloor(n+1)^2/4\rfloor-1}\right)
 =\Theta_q(N_q(n)q^{n/2}).
\end{equation}
Indeed, the summand has order
$q^{r(n-r)+n-r-1}=q^{(r+1)(n-r)-1}$. Completing the square gives
the displayed exponent, and terms farther from the maximum contribute a
bounded sum of powers of the form $q^{-j^2}$, as above.

\section{Running time for incidence between two dimensions}\label{app:incidence-work}
  The extra inside call in \eqref{eq:rank-work} repeats some entries.
We pay for it using the decrease of the input dimension or output
codimension in the frequency calls.
Put $c=n-b$, $A=|\Gr(n,a)|$, $B=|\Gr(n,b)|$, and $S=A+B$.
Write
  \[
 I=|\Gr(n-1,a)|+|\Gr(n-1,b)|,\qquad
 J=|\Gr(n-1,a)|+|\Gr(n-1,b-1)|
\]
  for the lengths of the two inside calls.
The fixed-dimension version of \eqref{eq:count} shows that the
frequency calls together have $S-I$ entries. All recursive calls
therefore have $S+J\le2S$ entries.

Assign a call the budget $\min(a,c)S$. Exchanging $a,c$ swaps the
first and third calls in \eqref{eq:rank-work}, so assume $a\le c$.
Every frequency call reduces this minimum by one. The first inside
call does so exactly when $a=c$, while the second keeps it unchanged.
Thus the budget decreases by
\begin{equation}\label{eq:slack}
 \min(a,c)S-\sum_{\text{recursive calls}}\min(a',c')S'
 =\begin{cases}S-aJ,&a=c,\\ S-I-aJ,&a<c,\end{cases}
 \quad\ge S/4.
\end{equation}
To verify the inequality, use
$|\Gr(n-1,j)|/|\Gr(n,j)|\le q^{-j}$.
If $a=c$, annihilators give $J\le q^{-a}S$, so the decrease is
at least $(1-aq^{-a})S\ge S/2$.
If $a<c$, then $a<b<n-a$, so $B\ge A$, and
\[
 S-I-aJ\ge\bigl(1-(a+1)q^{-a}\bigr)A
             +\bigl(1-q^{-b}-aq^{-c}\bigr)B
 \ge B/2\ge S/4.
\]
The last step uses $b,c\ge a+1$, $(a+1)q^{-a}\le1$, and
$(a+1)q^{-(a+1)}\le1/2$.

The boundary calls cost $O(S)$. Charge nonroot calls with $a=0$ or $c=0$
to their parents, using the bound $S+J\le2S$.
Since child dimensions decrease, multiplying \eqref{eq:slack} by $n$
shows that a constant multiple of $n\min(a,c)S$ pays for the local
$O_q(nS)$ work and all remaining calls.
For an interior call, with $h=b-a$, \eqref{eq:gaussian} gives
\[
 n\min(a,c)\le2\bigl(ac+h\max(a,c)\bigr)
 \le2\log_q S.
\]
This proves $O_q(S\log S)$. Charging $\poly(n)$ per entry in the
same argument also bounds index construction by $S\poly(n)$.

 \section{Further applications on the Grassmannian}\label{app:grassmann}
 We derive further consequences of \cref{thm:kernels}, using the same notation.

\paragraph{Distance sums.}
For $k=0,n$, the only distance sum is $f$. Otherwise first suppose
$k\le n/2$ and reuse the table $H$ from \eqref{eq:moments}.
To recover all distance sums, apply a lower transform to
$H(L)\ind[\dim L=j]$ for each $0\le j\le k$. This computes all
$m_j$ in \eqref{eq:moments}. For fixed $U$, let
$h_t(U)=\sum_{W\in V:\,\dim(U\cap W)=t}f(W)$, $0\le t\le k$.
Grouping \eqref{eq:moments} by intersection dimension gives
\begin{equation}\label{eq:profile}
 \begin{aligned}
 m_j(U)&=\sum_{t=j}^k|\Gr(t,j)|h_t(U),\\
 h_t(U)&=m_t(U)-\sum_{r=t+1}^k|\Gr(r,t)|h_r(U).
 \end{aligned}
\end{equation}
Back-substitution starts with $h_k=m_k$ and recovers
$h_{k-1},\ldots,h_0$, giving the desired outputs $D_if=h_{k-i}$.
The $k+2$ transforms and $O(k^2N)$ substitution work take
$O_q((k+1)N\log N)$ operations. Annihilators preserve distance and handle $k>n/2$, so the
full bound is $O_q((d+1)N\log N)$.

\paragraph{Spectral projections.}
To recover each eigenspace component of $f$, we construct a polynomial
that is one on its eigenvalue and zero on the others.
Over $\Q$ or $\C$, the real symmetric adjacency matrix $D_1$ has
eigenvalues~\cite{Sri14,MOR03}
\[
 \theta_j=q\qi k\qi{n-k}-\qi j\qi{n-j+1},\qquad 0\le j\le d.
\]
They are distinct because $\theta_j-\theta_{j+1}=q^j\qi{n-2j}>0$
for $j<d$. The polynomial and the $\theta_j$-component $f_j$ are
\[
 P_j(z)=\prod_{\substack{0\le t\le d\\t\ne j}}
            \frac{z-\theta_t}{\theta_j-\theta_t},\qquad
 f_j=P_j(D_1)f.
\]
Compute $f,D_1f,\ldots,D_1^{d}f$ by fast kernel multiplication,
then combine them using the coefficients of $P_j$. The total cost is
$O_q((d+1)N\log N)$, including the
$O((d+1)^3+(d+1)^2N)$ interpolation work.
For $d=0$ the projection is the identity.

\paragraph{Invariant linear systems.}
Over a field of characteristic different from $p$, let
$K(U,W)=a_{\operatorname{dist}(U,W)}$. To test invertibility and recover one inverse
column, fix $E\in V$ and put $\mathcal O_j=\{U:\operatorname{dist}(U,E)=j\}$.
Invertible linear maps fixing $E$ are transitive on each $\mathcal O_j$, by
extending bases, and commute with $K$. Thus $K$ preserves the span
of the indicators $b_j=\ind_{\mathcal O_j}$. Choosing $U_i\in\mathcal O_i$
gives its matrix on this space:
\begin{equation}\label{eq:small-system}
 B_{ij}=(Kb_j)(U_i),\qquad 0\le i,j\le d.
\end{equation}
If $K$ is invertible, its inverse also preserves this space, so $B$
is invertible. Conversely, if $B$ is invertible, solve
$Bc=(1,0,\ldots,0)^{\mathsf T}$ and
set $C(U,W)=c_{\operatorname{dist}(U,W)}$. Then $(KC)(U,E)=\ind[U=E]$, and invariance gives every other column, proving $KC=I$.
Constructing $B$, solving this small system, and applying $C=K^{-1}$
therefore test nonsingularity and solve $Kx=b$ in
$O_q((d+1)N\log N)$ operations.
No division by $|\mathcal O_j|$ is needed, even when eigenvalues
of $D_1$ coincide in the coefficient field.

\paragraph{Distances to a specified family of subspaces.}
For a nonempty family $\mathcal C\subseteq V$, also called a
constant-dimension code~\cite{KK08}, compute $D_i\ind_{\mathcal C}$ over
$\Q$. These values count codewords at each distance, so
\[
 \delta_{\mathcal C}(U)=\min\{i:(D_i\ind_{\mathcal C})(U)>0\}
\]
is the distance to the nearest codeword. The usual subspace distance
is twice the graph distance, giving covering radius $2\max_{U\in V}\delta_{\mathcal C}(U)$.
For $|\mathcal C|\ge2$, its minimum code distance is twice the
smallest $i>0$ for which $(D_i\ind_{\mathcal C})(U)>0$ for some $U\in\mathcal C$. Summing counts over $\mathcal C$ gives ordered-pair distance distributions.
Computing all distance sums gives these counts, but visits every subspace
in $V$ and need not be efficient for a sparse code or a single query.

\paragraph{Different input and output dimensions.}\label{sec:rectangular-kernels}
With $\mathcal R$ as in \cref{thm:linear}, let $K$ have columns indexed
by $\Gr(n,a)$ and rows by $\Gr(n,b)$, with $K(W,U)$ depending only on
$\dim(U\cap W)$. For input $f:\Gr(n,a)\to\mathcal R$, the same double count gives, for
$W\in\Gr(n,b)$ and $0\le t\le\min(a,b)$,
\begin{equation}\label{eq:rect-moments}
 \bigl(\IncidenceTransform_n^{t,b}(\IncidenceTransform_n^{t,a})^{\mathsf T}f\bigr)(W)
   =\sum_{U\in\Gr(n,a)}|\Gr(\dim(U\cap W),t)| f(U).
\end{equation}
If $a,b>n/2$, take annihilators,
which change the intersection dimension to $n-a-b+\dim(U\cap W)$.
Now at least one of $a,b$ is at most $n/2$, so each
intermediate table is no larger than one of the input and output tables.
Back-substitution in \eqref{eq:profile} recovers sums by intersection
dimension, or applies any intersection kernel, in $S\poly(n)$ work,
where $S=|\Gr(n,a)|+|\Gr(n,b)|$.
For unions of dimensions, sum over dimension pairs and let $S$ be
the combined table length. On the table of all subspaces, one upper transform
per input dimension and one weighted lower transform per output
dimension give $O_q((n+1)^3N_q(n))$ operations.

\section{Index construction and exact bit costs}\label{app:implementation}
\paragraph{Constructing the tables.}
To locate the entries used by each recursive transform,
represent subspaces by reduced row-echelon bases and cosets by reduced
representatives. Gaussian elimination constructs annihilators, kernels,
quotients, and coordinate changes. For a call on $\F_q^n$ and a
crossing direction $B\le\F_q^{n-1}$, enumerate $(B,\lambda)$ with
$\lambda\in B^\circ$. Express $B$ in a chosen basis of $\ker\lambda$
to locate its entry in the transform for frequency $\lambda$.
Group these entries by $\lambda$ for the smaller transforms and
by $B$ for Fourier inversion.
Each call has $q^{O(n^2)}$ records of $O(n^2)$ field elements, so elimination and sorting
cost $\poly(n)$ per record. Equations~\eqref{eq:count} and~\eqref{eq:slack}
charge the recursive records, including repeated incidence inputs,
giving table size times $\poly(n)$ construction work.

\paragraph{Exact integer and modular evaluation.}
\begin{cor}\label{cor:integer-lifting}
For fixed $q$, suppose a transform in \cref{thm:linear,thm:incidence}
has integer input of magnitude below $2^H$ and combined table length
$S$. After index construction, its exact output takes
\[
 O_q\!\left(S(H+n^2+1)^2\log S\right)
\]
bit operations. Reduction modulo any positive integer adds its reduction
cost, and scaling handles rational inputs with a supplied common denominator.
\end{cor}
\begin{proof}
We bound the outputs, then compute modulo a larger integer.
For dimension gap $j$, the inverse containment coefficient is
$(-1)^jq^{j(j-1)/2}$, as \eqref{eq:pascal} verifies by interval inversion. Cutoffs preserve intervals. Thus all transform coefficients are integers
of magnitude at most $q^{n(n-1)/2}$, and outputs have magnitude below
$B=2^H S q^{n(n-1)/2}$.
Choose $L\in\{2B+1,2B+2\}$ coprime to $p$.
The transform applies over $\mathbb Z/L\mathbb Z$, and its centered
residues uniquely recover the integer outputs because $L>2B$.
The modulus has $H+O_q(n^2+1)$ bits, so schoolbook modular arithmetic
gives the bound.
\end{proof}

For fixed, explicitly represented finite coefficient fields, evaluate
prime-field coordinates over the integers and reduce, even in
characteristic $p$, though the arithmetic bound still requires invertible $p$.
Span convolution has polynomial bit overhead, and exact-span counting uses
$H=\ell+1$.
For rational kernel coefficients of bit length $H$, determinant bounds
in \eqref{eq:small-system} give inverse coefficients of polynomial bit
length. Consequently kernel applications and solving take table size
times $\poly(n,H)$ bit operations for integer entries or entries sharing
a supplied common denominator of bounded bit length.

\section{Experimental details}\label{app:experiments}

The \texttt{experiments/} directory of the repository~\cite{PW26code}
contains the raw measurements, benchmark drivers, and analysis scripts.
Its README gives the commands for reproducing the tables and figure
from the measurements and for running the experiments again.

\paragraph{Implementation and measurements.}
We measured running times on one Intel Xeon Platinum 8273CL core
(2.2\,GHz), with double precision and $2$\,TB of available memory. The measurements exclude index construction.
The C++17 implementations use GCC~11.4 with \texttt{-O3 -march=native}
and share finite-field arithmetic and subspace indexing. Input weights
are uniform in $[-1,1]$ with fixed pseudorandom seeds.

For the zeta transform $\ZetaTransform_n$, we ran each algorithm three
times and kept its shortest measured running time. We repeated this
procedure $3$--$21$ times, changing which algorithm ran first each time.
We report the median of the retained measured running times and compute
each speedup from the two times in the same repetition. The figure shows
the median and quartiles of these speedups.
For the incidence transform and Grassmann kernels, our measured running
times are medians of $3$--$15$ and $3$--$25$ runs, respectively, after a warm-up.
Long incidence comparisons were run once. For repeated Poisson runs,
the measured running time is the sum of the medians for its three stages.
Operation counts use real arithmetic, including the operations needed for complex
Fourier arithmetic for odd $q$. Memory figures count the specified
tables, workspace, and indices in decimal GB or TB, not peak process memory.

\paragraph{Zeta transform.}
Our zeta transform uses the Fourier recursion without a covering-pair
base case. On $\F_2^{10}$, the covering-pair algorithm uses $0.97$ times
as many arithmetic operations as ours, but its measured running time is
$2.0$ times ours.
On $\F_{32}^5$, with $N_{32}(5)\approx2.2\cdot10^9$, our measured running
time is $145$\,s, whereas the covering-pair algorithm would need $5$\,TB
just to store one index per covering pair.
To estimate its running time, we processed about $2\%$ of the covering pairs from
$64$ evenly spaced ranges of output subspaces, using the full value
table, and scaled by the ratio of total to sampled pairs.
This estimate is $95$ times our measured running time, and the figure marks it
separately from measured running times. Applying the same procedure
on $\F_{16}^5$ and $\F_8^6$ underestimates their measured running times by
$11\%$ and $16\%$, respectively, but does not determine the error on
$\F_{32}^5$.
The operation count ratios in the figure are computed from the recursion
and exact covering-pair counts, including instances too large to run.

\paragraph{Incidence between separated dimensions.}
For $\IncidenceTransform_n^{a,b}$, direct summation processes
$|\Gr(n,b)|\,|\Gr(b,a)|$ incident pairs, while the covering-pair algorithm
passes through the intermediate dimensions, using
$\sum_{j=a+1}^b|\Gr(n,j)|\qi j$ additions.
Table~\ref{tab:exp-incidence} compares these operation counts with ours.

For Poisson summation, write $t=a=n-b$, as in all instances in
the table. Place $f(U)$ at the ordered reduced row-echelon basis of $U$ in a
table $F$ on $(\F_q^n)^t$, and set its other entries to zero.
Every $U\le W$ has exactly one such basis in $W^t$, so Poisson
summation~%
\cite[Theorem~10.1.16]{MP13} %
gives
\[
 (\IncidenceTransform_n^{t,n-t}f)(W)
 =\sum_{x\in W^t}F(x)
 =q^{-t^2}\sum_{y\in(W^\circ)^t}\widehat F(y).
\]
The Fourier table has $q^{tn}=\Theta_q(q^{t^2}S)$ entries, by
\eqref{eq:gaussian}.
For $q=p^m$, the table's ``FFT bound'' is the conservative count
$tmnq^{tn}$, charging one operation per entry per base-$p$ coordinate.
It omits the final sums and undercounts the real arithmetic for odd
$p$. The measured running time includes constructing $F$, taking its Fourier transform,
and summing the required frequencies.

\begin{table}[tbp]

\centering\small
\setlength{\tabcolsep}{4pt}
\begin{tabular}{@{}rlrrrrrrr@{}}
\toprule
&&& ours & \multicolumn{3}{c}{operation count / ours} & \multicolumn{2}{c}{time / ours}\\
\cmidrule(lr){4-4}
\cmidrule(lr){5-7}
\cmidrule(l){8-9}
$q$ & $(n,a,b)$ & $S$ & time & direct & covering & FFT bound & covering & Poisson\\
\midrule
2 & $(16,1,15)$ & $1.3\cdot10^5$ & $1.9$\,ms & $1600$ & $3.0\cdot10^{16}$ & $0.76$ & -- & $0.68$\\
2 & $(14,2,12)$ & $8.9\cdot10^7$ & $2.1$\,s & $7.1\cdot10^4$ & $3.5\cdot10^8$ & $4.3$ & -- & $4.5$\\
3 & $(8,2,6)^\dagger$ & $1.8\cdot10^6$ & $32$\,ms & $120$ & $82$ & $8.4$ & $550$ & $42$\\
4 & $(8,2,6)^\dagger$ & $4.8\cdot10^7$ & $1.0$\,s & $2700$ & $1400$ & $170$ & -- & $130$\\
4 & $(9,2,7)$ & $7.6\cdot10^8$ & $20$\,s & $3.9\cdot10^4$ & $5.6\cdot10^4$ & $170$ & -- & $190$\\
8 & $(6,2,4)$ & $3.9\cdot10^7$ & $810$\,ms & $140$ & $34$ & $3700$ & $140$ & $5300$\\
8 & $(7,2,5)$ & $2.5\cdot10^9$ & $80$\,s & $7500$ & $1100$ & $3600$ & -- & --\\
2 & $(10,3,7)$ & $1.3\cdot10^7$ & $260$\,ms & $300$ & $33$ & $130$ & $84$ & $140$\\
2 & $(11,3,8)$ & $1.0\cdot10^8$ & $2.9$\,s & $2200$ & $210$ & $130$ & -- & $130$\\
2 & $(12,3,9)^\dagger$ & $8.2\cdot10^8$ & $28$\,s & $1.6\cdot10^4$ & $1900$ & $130$ & -- & $160$\\
\bottomrule
\end{tabular}
\caption{Incidence transforms $\IncidenceTransform_n^{a,b}$ with $b=n-a$
 and $S=|\Gr(n,a)|+|\Gr(n,b)|$. Ratios above one favor our algorithm.
 Operation counts are divided by our count of real arithmetic
 operations. ``FFT bound'' is the conservative count for Poisson
 summation described in the text. A dash means the comparison was
 not run because of memory or indexing limits: the
 covering-pair runs allow $400$\,GB of estimated memory and 32-bit
 table indices. ${}^\dagger$ marks the family $(4r,r,3r)$ of
 \cref{thm:optimal}.}
\label{tab:exp-incidence}

\end{table}

For $\IncidenceTransform_7^{2,5}$ over $\F_8$, our tables, workspace,
and indices use about $70$\,GB. The intermediate layers alone require
$1.3$\,TB of values before storing any covering pairs, while the
Fourier table for Poisson summation requires $35$\,TB, so neither comparison algorithm fits in memory.
When $t=1$, the Poisson table has only $q^n$ entries and each final
sum has $q$ terms, explaining its shorter measured running time in the first row.

\paragraph{Grassmann kernels.}
For $L\in\Gr(n,k-1)$ and $U\in\Gr(n,k)$, let $B(L,U)=\ind[L\le U]$.
The identity~%
\cite[Section~5.3]{MOR03}
\[
 D_1=B^{\mathsf T}B-\qi k I
\]
reduces each adjacency multiplication to two incidence sums, with
$2\qi k|\Gr(n,k)|$ additions and a diagonal correction, without
storing the adjacency matrix.
The recurrence uses real arithmetic even for odd $q$, and we conservatively
count only its $2d\qi kN$ incidence additions, omitting the scalar work
in the recurrence and weighted sum.
Our count includes both transforms, the intermediate scalings, and the
coefficient calculation. A complex scaling counts as two real multiplications.

\begin{table}[tbp]

\centering\small
\begin{tabular}{@{}lrrcc@{}}
\toprule
& & ours & \multicolumn{2}{c}{time / ours}\\
\cmidrule(l){4-5}
Graph & $N=|\Gr(n,k)|$ & time & \shortstack{distance-matrix\\recurrence} & \shortstack{two covering-pair\\transforms}\\
\midrule
$J_2(10,5)$ & $1.1\cdot10^8$ & $17$\,s & $6.2$ & $1.23$\\
$J_3(8,4)$ & $7.6\cdot10^7$ & $10$\,s & $8.0$ & --\\
$J_5(6,3)$ & $2.6\cdot10^6$ & $170$\,ms & $4.8$ & --\\
$J_8(6,3)$ & $1.6\cdot10^8$ & $17$\,s & $22$ & $6.6$\\
\bottomrule
\end{tabular}
\caption{Large Grassmann kernel instances, with $a_i=2^{-i}$.
 The last column replaces the fast transforms in
 \eqref{eq:two-transforms} by covering-pair transforms and is reported
 only in characteristic two. Measured running times exclude index construction.}
\label{tab:exp-kernels}

\end{table}

\paragraph{Checking the outputs.}
We check direct sums on all outputs of small instances and sampled
outputs of large ones ($200$ per incidence instance), and compare all
outputs whenever both methods produced all outputs. For computed values $y$,
reference values $y_{\mathrm{ref}}$, and each set $T$ of checked outputs,
the relative discrepancy satisfies
\[
 \frac{\max_{W\in T}|y(W)-y_{\mathrm{ref}}(W)|}{\max_{W\in T}|y_{\mathrm{ref}}(W)|}<10^{-13}.
\]
These checks do not give a numerical-stability guarantee. Exact arithmetic is covered by
\cref{sec:arithmetic,app:implementation}.

\section{A matching lower bound for incidence}\label{sec:lower-bound}
\begin{thm}\label{thm:optimal}
For $r\ge1$ and $v=|\Gr(4r,r)|$, applying $\IncidenceTransform_{4r}^{r,3r}$ over $\C$
requires $\Omega_q(v\log v)$ gates of the form
$z=\alpha x+\beta y$, $|\alpha|,|\beta|\le1$.
The algorithm of \cref{thm:incidence} attains this bound up to a
constant factor depending on $q$.
\end{thm}
\begin{proof}
We lower-bound the determinant and apply Morgenstern's
bounded-coefficient argument~\cite{Mor73}.
  Put $I=\IncidenceTransform_{4r}^{r,3r}$. For $r$-spaces $U,W$ with
$t=\dim(U\cap W)$, quotienting by $U+W$ gives
\[
 (I^{\mathsf T}I)(U,W)=|\Gr(2r+t,r+t)|=|\Gr(2r+t,r)|.
\]
  Count the $r$-spaces in $\F_q^{2r}\oplus(U\cap W)$ by their
projection $L\le U\cap W$. For a fixed $j$-dimensional $L$, choose
an $(r-j)$-space $A\le\F_q^{2r}$ and a map $L\to\F_q^{2r}/A$.
There are $c_j=|\Gr(2r,r-j)|q^{j(r+j)}$ choices. Thus
$(I^{\mathsf T}I)(U,W)=\sum_{L\le U\cap W}c_{\dim L}$, and every real
vector $x=(x_U)_{U\in\Gr(4r,r)}$ satisfies
\[
 \|Ix\|_2^2
 =\sum_{\substack{L\le\F_q^{4r}\\\dim L\le r}}
       c_{\dim L}\left(\sum_{\substack{U\in\Gr(4r,r)\\L\le U}}x_U\right)^2
 \ge c_r\|x\|_2^2=q^{2r^2}\|x\|_2^2.
\]
 The inequality keeps just the terms with $\dim L=r$.
Every singular value of the square matrix $I$ is therefore at least
$q^{r^2}$. Since $\log v=\Theta_q(r^2)$ by \eqref{eq:gaussian},
$\log|\det I|\ge vr^2\log q=\Omega_q(v\log v)$.

 Regard each wire as a row of coefficients in the $v$ inputs. Initially
every $v$-row determinant has absolute value at most one. A gate
$\alpha x+\beta y$ with $|\alpha|,|\beta|\le1$ increases the largest
such determinant by at most a factor of two, by multilinearity.
Thus at least $\log_2|\det I|$ gates are needed.
The upper algorithm
uses only additions, roots of unity, and Fourier normalizations of
modulus at most one, so it fits this model. This is a lower bound for
the stated incidence matrices, not for unrestricted constants or for
the full zeta matrix, whose determinant is one.
\end{proof}

 \section{The counter reduction}
\label{xp:section}\label{xp:overview}

We adapt the monotone-counter reduction of~\cite{FGL+19}, replacing
binary variable identifiers by subspaces of $\F_2^{2d}$. A constraint
uses colors of a specified type, and a single clique will test all these
colors together.

\begin{thm}\label{xp:lower}
Assuming $\ETH$, for every computable function $f$ and fixed $\alpha\ge0$,
there is no algorithm deciding whether $\chi(G)\le\eta$ in time
\[
 f(r)N^{\,\alpha+2^{r^2/4-17r}}
\]
on $N$-vertex graphs supplied with a linear layout of width $r$, where
$\eta$ is part of the input. The same exclusion holds with a supplied
rank-decomposition of width $r$.
\end{thm}

\subsection{Monotone counters and the register}

An instance of \textsc{4-Monotone min-CSP} has $k$ variables with
values in $\{0,\ldots,n\}$, a budget $W$, and constraints $\mathcal C$.
A constraint $C$ involves a set $X_C$ of at most four variables and a
nonempty family $\mathcal R_C$ of \emph{mini-constraints}.
Each $R\in\mathcal R_C$ requires $\alpha(x)\ge c_{R,x}$ for every
$x\in X_C$, where $c_{R,x}\in\{0,\ldots,n\}$.
We seek an assignment of cost $\sum_x\alpha(x)\le W$ satisfying
at least one mini-constraint of each $C$. Increasing values preserves satisfaction.

  \begin{lem}[\cite{FGL+19}, Lemma~3.1 and its construction] \label{xp:source}
Under $\ETH$, for any computable $f$, this problem has no
$f(k)n^{o(k)}$ algorithm, even when
$|\mathcal C|=O(k^2n^2)$ and
$L:=\max(1,\max_C|\mathcal R_C|)\le\max(n,4)$.
\end{lem}

The construction in~\cite[Lemma~3.1]{FGL+19}
uses $2K$ counters for a $K$-part
 \textsc{Multicolored Clique} instance with $n$ vertices per part.
It has $K$ constraints with $n$ mini-constraints and at most
$\binom K2n^2$ constraints with at most four mini-constraints.
Discard mini-constraints with thresholds greater than $n$ and pad missing
inequalities with threshold zero. Empty families give an
immediate rejection, while empty scopes are tautologies and can be deleted.
We may handle $k<2$, $n=0$, and budgets outside $[0,kn]$ directly.

Put $d=\lceil\sqrt{\log_2 k}\rceil$, $M=2^{d^2}$, $h=2^d$, and $Q=Mn$.
Pad with unused variables and index the resulting $M$ variables by
matrices $A\in\F_2^{d\times d}$. Make $d$ disjoint $Q$-cliques $B_i$,
  with $Q/h$ vertices of each label $(e_i,a)$, $a\in\F_2^d$, in $B_i$.
Also make a disjoint clique $B_*$ of size $Q-W$.
The construction below forces all rows to use a common $Q$-color palette
containing the colors of $B_*$. Each color occurs once in each row, and its labels $(e_i,A^{\mathsf T}e_i)$ define its matrix type $A$.
 The $W$ colors absent from $B_*$ are called \emph{active}, and their number
of type $A$ represents $x_A$.

  To identify a color's type, associate $A$ with the subspace
$S_A=\{(x,A^{\mathsf T}x):x\in\F_2^d\}$ and the basis
$t_{A,j}=(Ae_j,e_j)$, $1\le j\le d$, of $S_A^\perp$.
A color of type $T$ passes every orthogonality test exactly when $T=A$, since
\begin{equation}\label{xp:entry-test}
 (e_i,T^{\mathsf T}e_i)^{\mathsf T}t_{A,j}=T_{ij}+A_{ij}.
\end{equation}

Every family of integers $0\le\alpha_A\le n$ with $\sum_A\alpha_A=W$
  can be realized as the active color counts. Allocate $\alpha_A$ distinct
colors to type $A$ and place them on its label in every row. Exactly
$M/h$ matrices have a prescribed row, so no label receives more than
 $(M/h)n=Q/h$ colors. In each row independently, fill its remaining
 $Q-W$ vertices bijectively with the same $Q-W$ new colors, and use those
on $B_*$. There are no edges between rows. A satisfying assignment of
cost below $W$ can be increased to cost $W$: the total capacity is
$Mn\ge W$, and increasing values preserves every source constraint.

\subsection{Constraint gadgets}

  For each threshold $x_A\ge c$, an $n$-clique will need at least $c$
 active colors of type $A$. A selector chooses a mini-constraint and
makes at most $n-c$ other colors available to this clique.
 For each $C$ and $x\in X_C$, make an $n$-clique $A_{C,x}$.
For each $R\in\mathcal R_C$ and $x\in X_C$, make an independent set
$I_{C,R,x}$ of size $n-c_{R,x}$. Write $I_{C,R}=\bigcup_x I_{C,R,x}$,
$I_{C,x}=\bigcup_R I_{C,R,x}$, and $I_C=\bigcup_R I_{C,R}$.
Join distinct $I_{C,R}$ completely for fixed $C$, and join every
 $I_{C,R,x}$ completely to $A_{C,x}$.
There are no other edges among these $A$- and $I$-vertices.

  Make a selector $s_C$ for each $C$ and a private anchor $m_b$ for each
$I$-vertex $b$. All selectors and anchors form a clique $D$, completely
joined to every $B_i$ and to $B_*$.
  Join $s_C$ to every $A$- and $I$-vertex outside $I_C$.
If $b\in I_{C,R,x}$, join $m_b$ to every $A$- and $I$-vertex outside
$A_{C,x}\cup\{b\}$. Denote the anchors belonging to $I_{C,x}$ by
$\mathcal M_{C,x}=\{m_b:b\in I_{C,x}\}$.
Join all $A$- and $I$-vertices to $B_*$, and all $I$-vertices to every
$B_i$. Set the target number of colors to
\begin{equation}\label{xp:quota}
 \eta=Q+|D|.
\end{equation}
  In an $\eta$-coloring, the $\eta$-cliques
$B_i\cup D$ force each $B_i$ to use exactly the complement of the
 $D$-colors. Since $B_*$ is complete to $D$, its colors lie in this
common register palette.

It remains to restrict the register colors on $A_{C,x_A}$ to type $A$.
  The test needs to copy only the set of colors on this clique.
Call the vertices constructed so far \emph{old vertices}. The following
cliques have only the edges specified below.
For each $A_{C,x_A}$, make $d$ $n$-cliques
$A^{(1)}=A_{C,x_A},A^{(2)},\ldots,A^{(d)}$, of which only the first keeps
the edges already specified. Join every vertex of $A^{(j)}$ to a
register vertex of label $v$ exactly when
$v^{\mathsf T}t_{A,j}=1$. Add a private $(\eta-n)$-clique
  $E_{C,x_A}$, complete to these $d$ cliques and with no other neighbors.
These test cliques together with $E_{C,x_A}$ form a \emph{palette block}.

  The private clique leaves exactly $n$ colors, so every $A^{(j)}$ uses
the same palette. Each register color in this palette passes all the
tests in~\eqref{xp:entry-test}, and therefore has type $A$.
Conversely, any coloring satisfying the old edges and using only type-$A$
register colors on $A^{(1)}$ extends: use its palette on every test clique and all other colors on
$E_{C,x_A}$. Colors from $D$ have no register conflicts.
After fixing the register and $D$, identify anchors with their colors
and let $P_A$ be the active colors of type $A$. The remaining problem
is exactly list coloring of the old $A/I$ graph with
\[
 L(b)=\{s_C,m_b\}\quad(b\in I_C),\qquad
 L(a)=P_A\cup\mathcal M_{C,x_A}\quad(a\in A_{C,x_A}).
\]
The specified edges give precisely these lists, and every such coloring
extends independently to the palette blocks.

\subsection{Correctness}

\begin{lem}\label{xp:correct}
The graph is $\eta$-colorable if and only if the source instance is satisfiable.
\end{lem}
\begin{proof}
  Given an $\eta$-coloring, let $z_A$ count the active colors of type $A$, so $\sum_Az_A=W$. Since distinct $I_{C,R}$ are completely joined,
at most one mini-constraint $R$ has a vertex using the selector color.
Choose that $R$, or any $R$ if none does. Every vertex belonging to
another mini-constraint uses its private anchor color. Its complete join
to $A_{C,x}$ blocks that color on the whole clique. For $x_A\in X_C$,
  at most $|I_{C,R,x_A}|=n-c_{R,x_A}$ anchor colors therefore remain
available to $A_{C,x_A}$. Its $n$ vertices have distinct colors, so
\[
 n\le(n-c_{R,x_A})+z_A,\qquad z_A\ge c_{R,x_A}.
\]
Consequently $\alpha(x_A)=\min(n,z_A)$ satisfies every constraint
and has total cost at most $W$. Restrict it to the original variables.

  Conversely, increase a satisfying padded assignment to cost $W$ and
realize it in the register, using colors disjoint from those of $D$.
  Choose a satisfied mini-constraint $R$ for each $C$. Color $I_{C,R}$ with
the selector color and every other $b$ with its private color.
Color $A_{C,x_A}$ with the $n-c_{R,x_A}$ private colors from
$I_{C,R,x_A}$ and $c_{R,x_A}$ distinct active colors of type $A$.
This is a proper list coloring: different $I$-groups use disjoint
colors, and $A$ uses no unchosen group's private color. Extend the
palette blocks as above. Empty $I$-sets cause no change.
 \end{proof}

\subsection{The linear layout}

\begin{lem}\label{xp:width}
The construction supplies a linear layout of cut-rank at most $2d+12$.
\end{lem}
\begin{proof}
Order the row cliques first, then $B_*$. Process constraints consecutively.
For each $C$, put $s_C$ first, then its palette blocks one at a time,
then its mini-constraints one at a time. Within a palette block, order
the whole cliques $A^{(1)},\ldots,A^{(d)},E_{C,x}$ consecutively.
Within a mini-constraint, put each pair $m_b,b$ consecutively.

  The register-to-test edges have rank at most $2d$: they factor through
the register labels in $\F_2^{2d}$, and all test vertices follow the
register. The universal columns to $D$ and $I$ lie in the same space,
since the sum of a register label's first $d$ coordinates is one.
During the register prefix at most one row clique is split, and during
 $B_*$ its processed vertices have one common future neighborhood.
These initial cuts therefore have rank at most $2d+1$.

Consider a later cut, in the part for constraint $C$. Remove the
register edges, the edges within test cliques, and the private-clique
edges. The remaining nonzero rows have at most ten different future
neighborhoods:
\begin{itemize}
 \item $D$-vertices whose exceptions are in the past, and $B_*$, see
       all future old vertices, whereas $A/I$-vertices of completed constraints
       see exactly future $D$-vertices.
   \item The current selector gives one further neighborhood.
 \item Current $A$-vertices give at most four, one for each $x\in X_C$:
       they see future $D$ outside $\mathcal M_{C,x}$ and future $I_{C,x}$.
 \item Current $I$-vertices give at most two, according to whether
       their mini-constraint is current or already finished.
  \item A split pair $m_b,b$ adds at most one exceptional anchor.
\end{itemize}
Indeed, all $A$-cliques precede the anchors, and each finished anchor
has both its $A$-clique and its exceptional vertex $b$ in the past.
Thus the remaining rank is at most $2+1+4+2+1=10$.
At most one test clique is split, contributing rank one. At most
  one palette block is split, and its crossing private edges form a complete
bipartite graph, contributing rank one. Adding back the register edges
gives cut-rank at most $2d+10+1+1$.
\end{proof}

\subsection{Size and the ETH exponent}

\begin{proof}[Proof of \cref{xp:lower}]
  Write $a=\sum_C|X_C|$ and $p=|\mathcal C|$. There are $a$
palette blocks, each with $dn+\eta-n$ vertices, and
$|D|=O(p+anL)$ anchors, so
\[
 N=O\bigl(dMn+p+anL+a(\eta+dn)\bigr)\le g(k)n^6
\]
  for a computable $g$: use $p=O(k^2n^2)$, $a\le4p$, and $L\le4n$
from \cref{xp:source}. Constructing all edges and the layout takes
$g_1(k)n^{O(1)}$ time with an absolute polynomial degree.
Put $s=\sqrt{\log_2k}$. The actual layout width satisfies
$0\le r\le2\lceil s\rceil+12\le2s+14$.
For the convex function $b(t)=t^2/4-17t$,
\[
 2^{b(r)}\le\max\{1,2^{b(2s+14)}\}
 =\max\{1,k2^{-20s-189}\}=o(k).
\]
  Bound $f(r)$ by the maximum of $1,f(j)$ over the integers
$0\le j\le2\lceil s\rceil+12$. Combining the hypothetical running
time with the construction cost and absorbing all factors depending
only on $k$ gives $F(k)n^{o(k)}$, contrary to \cref{xp:source}.
Finally, the graph contains a $Q$-clique with $Q\ge2$, so $r\ge1$.
The layout's caterpillar rank-decomposition has the same nonleaf cuts
and singleton cuts of rank at most one, and therefore the same width.
 \end{proof}

\section{Fixed-palette decision and counting lower bounds}
\label{fix:implementation}\label{fix:section} \label[appendix]{cnt:section}\label[appendix]{gad:section}

 \begin{thm}\label{thm:fixed-palette}
Assuming $\ETH$, for every fixed $c\ge3$, deciding $c$-colorability
has no $2^{o(r^2)}N^{O(1)}$-time algorithm on $N$-vertex graphs
supplied with a linear layout of width at most $r$. The same exclusion
holds for supplied rank-decompositions.
\end{thm}
 \begin{thm}\label{cnt:lower}
Assume $\#\ETH$. For every fixed $c\ge3$, exact counting of proper
$c$-colorings has no deterministic $2^{o(r^2)}N^{O(1)}$ algorithm,
even on graphs supplied with a linear layout of width at most $r$.
\end{thm}

One reduction proves both statements: each satisfying assignment
will give the same known number of labeled three-colorings.
Here $\#\ETH$ excludes $2^{o(n)}\poly(m)$ time for counting satisfying
assignments of $3$-CNF formulas with $n$ variables and $m$ clauses~\cite{DHM+14}.
Let $\varphi$ have $k^2$ variables in $k$ rows of length $k$, with
$k\ge1$, and $m$ clauses. Repeat literals in shorter nonempty clauses
so each clause has three literal occurrences. Write $J_i$ for the set
of rows containing variables of clause $i$.

  \paragraph{Making the row assignments agree.}
Create a triangle with vertices and colors named $T,F_A,F_B$.
An input is \emph{selected} when it has color $T$.
For every row $j$, create a base group of inputs $a_{0,j,u}$,
one for each assignment $u\in\F_2^k$ to that row. For each clause
$i$ and $j\in J_i$, create another group $a_{i,j,u}$ with the same
possible assignments. All these inputs have label
$(e_j,u)\in\F_2^{2k}$ and are adjacent to $F_B$, restricting their
colors to $\{T,F_A\}$.

  To make different copies agree, introduce one group of inputs
$b_{h,v}$, $v\in\F_2^k$, for each column $h\in[k]$.
Give these inputs labels $(v,e_h)$ and join them to $F_A$, restricting
their colors to $\{T,F_B\}$. Join an $A$ input to a $B$ input when
their labels have dot product one. Since their color sets intersect
only in $T$, selected labels must satisfy
\[
 (e_j,u)^{\mathsf T}(v,e_h)=0
 \quad\Longleftrightarrow\quad u_h=v_j.
\]
Suppose every group selects at least one input. A selected vector
$v_h$ in each column group fixes all coordinates of every selected
row assignment: $(u_j)_h=(v_h)_j$. Thus each row group selects
exactly one assignment, identically in all its copies. The base row
groups in turn fix every selected column assignment, so each column
group also selects exactly one input. Conversely, any matrix of row
assignments $u_j$ gives exactly these selections, with the $v_h$ as
its columns. Base groups ensure that variables absent from all clauses
are still represented.

  \paragraph{Enforcing a nonempty choice.}
The preceding argument already excludes multiple selections once
each group selects at least one input. An odd cycle suffices to enforce this
remaining condition. For a group $x_1,\ldots,x_t$, where $t=2^k$,
let $F$ be its false color, either $F_A$ or $F_B$. Add a fresh cycle
$w_0w_1\cdots w_tw_0$, join $w_0$ to $F$, and join $w_i$ to $x_i$
for $1\le i\le t$.

  If every input is false, all cycle vertices avoid $F$, which would
two-color an odd cycle. If exactly $x_j$ is selected, every cycle
vertex except $w_j$ still avoids $F$. Hence $w_j$ must have color
$F$, and the remaining path alternates the other two colors in
exactly two ways. Thus every coloring selects at least one input in each group,
 and every consistent singleton selection has exactly two extensions
on its cycle. No claim about larger selections is needed.

  \paragraph{Testing a clause with a fixed number of extensions.}
For each literal occurrence, make an adjacent pair $c,\bar c$, both
adjacent to $F_A$, so they have opposite colors in $\{T,F_B\}$.
  Join $c$ to the clause-local row inputs that falsify the literal,
and $\bar c$ to those that satisfy it. The unique selected input
then forces $c=T$ precisely when the literal is true, while unselected
inputs have color $F_A$ and impose no further restriction.

  For three terminals in $\{T,F_B\}$, a fresh triangle with one vertex
adjacent to each terminal enforces the not-all-equal relation
$\operatorname{NAE}$. Equal terminal colors leave only two colors
for the triangle. For mixed colors, the vertex opposite the singleton
terminal color must use the repeated color, and the other two vertices
can use the remaining colors in either order. Thus every valid input
has exactly two extensions, also for repeated terminals or constants.

  For a clause's three literal vertices $a,b,c$, introduce an adjacent pair
$s,\bar s$, both adjacent to $F_A$, and four private NAE triangles
imposing
\begin{equation}\label{gad:or}
\begin{aligned}
 &\operatorname{NAE}(\bar a,s,F_B),\qquad
   \operatorname{NAE}(\bar b,s,F_B),\\
 &\operatorname{NAE}(a,b,\bar s),\qquad
   \operatorname{NAE}(s,c,F_B).
\end{aligned}
\end{equation}
  The first two conditions say that $a$ or $b$ being true forces $s$
  true. The third forces $s$ false when both are false, and the last
requires $s\lor c$. Thus $s=a\lor b$ is uniquely determined,
and the clause has exactly $2^4=16$ extensions when satisfied and
none otherwise. All auxiliary vertices of the cycles and triangles are private, and there are no edges other than those specified.

  Every proper coloring now determines a satisfying assignment.
Conversely, a satisfying assignment fixes all inputs, literal pairs,
and auxiliary pairs once the palette colors are fixed. There are
$g=2k+\sum_i|J_i|$ groups, each with two cycle extensions, and each
clause contributes sixteen extensions. Therefore
\begin{equation}\label{cnt:identity}
 \#3\text{-Color}(G_\varphi)
   =6\cdot2^g\cdot16^m\,\#\mathrm{SAT}(\varphi),
\end{equation}
  where six counts the labeled palette colorings.

  \paragraph{The supplied layout and its size.}
Order the palette first, then the $B$ groups, then the base $A$
groups, then each clause's local $A$ groups followed by its clause
test. Within a group, use the order
$w_0,x_1,w_1,\ldots,x_t,w_t$. Within a clause test, put the four
literal and auxiliary pairs first and then the four triangles, one at a time.

  For any prefix cut, the $AB$ block has rank at most $2k$ by its
label factorization: all $B$ inputs precede all $A$ inputs.
Palette edges have rank at most three. Edges from row inputs to
literal pairs have at most six future columns, all in one clause.
For the remaining local edges, only one group or clause test can
straddle the cut. In a group, at most $w_0$, the last cycle vertex,
and the current input have later local neighbors. In a clause test,
at most eight vertices in the literal and auxiliary pairs and two vertices of the unfinished
triangle have later neighbors. Subadditivity therefore bounds every
cut-rank by $2k+3+6+10=2k+19$, including cuts inside pairs and
triangles.

  The graph has $3+g(2^{k+1}+1)+20m$ vertices: a group contributes
$2^k$ inputs and $2^k+1$ cycle vertices, and a clause contributes
eight vertices in the literal and auxiliary pairs and twelve triangle vertices.
Its $O((k+m)k2^{2k})$ edges and the layout are constructible in
 $2^{O(k)}\poly(m)$ time.
For an $n$-variable formula, take $k=\lceil\sqrt n\rceil$ and pad
to $k^2=n+O(\sqrt n)$ variables. For counting, divide out both the
factor in \eqref{cnt:identity} and the $2^{k^2-n}$ assignments to
the added variables. Empty clauses, empty formulas, and bounded
cases are handled directly. A $2^{o(r^2)}N^{O(1)}$ algorithm would
therefore decide or count SAT in $2^{o(n)}\poly(m)$ time, contrary
to $\ETH$ or $\#\ETH$~\cite{IP01,DHM+14}. The count and divisors
have $O(|V(G_\varphi)|)$ bits, so exact division fits this bound.

 For fixed $c>3$, add a universal $(c-3)$-clique and place it first.
It consumes exactly $c-3$ colors, adds at most $c-3$ to the layout
width, and multiplies the labeled count by $c!/3!$.
Finally, a linear layout gives a caterpillar rank-decomposition with
the same nonleaf cuts and leaf cuts of rank at most one. This proves
both theorems.

\label{page:appendix-end}
\paragraph{AI methodology statement.}
ChatGPT and Claude assisted with literature checks, proofs,
editing, and verification code for the transforms, linear algebra,
and coloring reductions. Claude Code also assisted with the Lean~4
formalization described in the introduction. The authors are responsible for the claims
and attribution. These checks are not external expert review.

\label{page:nonreference-end}
\clearpage
\begingroup
\setlength{\emergencystretch}{4em}
\raggedright
\printbibliography
\endgroup
\end{document}